\documentclass[10pt]{IEEEtran}
\usepackage{cite}
\usepackage{bm}
\usepackage{amsmath}
\usepackage{extarrows}
\usepackage{amssymb}
\usepackage{graphicx}
\usepackage{color}
\graphicspath{{figure/}}

\newcommand{\mb}{\bm}
\newcommand{\mr}{\mathrm}
\newcommand{\ms}{\mathrm}
\newcommand{\BE}{\begin{equation}}
\newcommand{\EE}{\end{equation}}
\newcommand{\BS}{\begin{subequations}}
\newcommand{\ES}{\end{subequations}}
\renewcommand{\bf}{\bm}

\newtheorem{theorem}{Theorem}
\newtheorem{proposition}{Proposition}
\newtheorem{assumption}{Assumption}
\newtheorem{definition}{Definition}

\newtheorem{lemma}{Lemma}

\begin{document}
\title{Orthogonal AMP}

\author{Junjie~Ma
        and~Li~Ping,~\IEEEmembership{Fellow,~IEEE}
          \thanks{          The work described in this paper was jointly supported by grants from University Grants Committee of the Hong Kong Special Administrative Region, China (Project numbers AoE/E-02/08, CityU 11217515 and CityU 11280216). }
}
\maketitle
\begin{abstract}
Approximate message passing (AMP) is a low-cost iterative signal recovery algorithm for linear system models. When the system transform matrix has independent identically distributed (IID) Gaussian entries, the performance of AMP can be asymptotically characterized by a simple scalar recursion called state evolution (SE). However, SE may become unreliable for other matrix ensembles, especially for ill-conditioned ones. This imposes limits on the applications of AMP. \par
In this paper, we propose an orthogonal AMP (OAMP) algorithm based on de-correlated linear estimation (LE) and divergence-free non-linear estimation (NLE). The Onsager term in standard AMP vanishes as a result of the divergence-free constraint on NLE. We develop an SE procedure for OAMP and show numerically that the SE for OAMP is accurate for general unitarily-invariant matrices, including IID Gaussian matrices and partial orthogonal matrices. We further derive optimized options for OAMP and show that the corresponding SE fixed point coincides with the optimal performance obtained via the replica method. Our numerical results demonstrate that OAMP can be advantageous over AMP, especially for ill-conditioned matrices.
\end{abstract}


\IEEEpeerreviewmaketitle

\section{Introduction}
Consider the signal recovery problem for the following linear model:
\BS  \label{Eqn:Intro}
\begin{align}
{\bf{y}} &= {\bf{Ax}} + {\bf{n}}, \label{Eqn:Intro_a}\\
x_j&\sim P_X(x),\quad \forall j,  \label{Eqn:Intro_b}
\end{align}
\ES
where $\bm{A}\in\mathbb{R}^{M\times N}$ ($M\le N$) is a channel matrix (for communication applications) or a sensing matrix (for compressed sensing), $\bm{x}\in\mathbb{R}^{N\times1}$ the signal to be recovered and $\bm{n}\in\mathbb{R}^{M\times1}$ is a vector of additive white Gaussian noise (AWGN) samples with zero mean and variance $\sigma^2$, and $P_X(x)$ a probability distribution with $\mr{E}\{x_j\}=0$ and $\mr{E}\{x_j^2\}=1$. We assume that $\{x_j\}$ are independent identically distributed (IID). Our focus is on systems with large $M$ and $N$.\par
Except when $P_X(x)$ is Gaussian or for very small $M$ and $N$, finding the optimal solution to \eqref{Eqn:Intro} (under, e.g., the minimum mean-squared error (MMSE) criterion \cite{Kay1993}) can be computationally prohibitive. Approximate message passing (AMP) \cite{Donoho2009} offers a computationally tractable option. AMP involves the iteration between two modules: one for linear estimation (LE) based on \eqref{Eqn:Intro_a} and the other for symbol-by-symbol non-linear estimation (NLE) based on \eqref{Eqn:Intro_b}. An Onsager term is introduced to regulate the correlation problem during iterative processing.\par
When $\bm{A}$ contains zero-mean IID Gaussian (or sub-Gaussian) entries, the dynamical behavior of AMP can be characterized by a simple scalar recursion, referred to as \textit{state evolution} (SE) \cite{Donoho2009,Bayati2011,Bayati2015}. The latter bears similarity to density evolution \cite{Richardson2001} (including EXIT analysis \cite{Brink2001}) for message passing decoding algorithms. However, the underlying assumptions are different: density evolution requires sparsity in $\bm{A}$ \cite{Richardson2001} while SE does not \cite{Bayati2011}. When $\bm{A}$ is IID Gaussian, it is shown in \cite{DonohoITW} that the fixed-point equation of the SE for AMP coincides with that of the MMSE performance for a large system. (The latter can be obtained using the replica method \cite{Guo2005Replica,Rangan2009,Tulino2013,Wen2014a}.) This implies that, when $\bm{A}$ is IID Gaussian, AMP is Bayes-optimal provided that the fixed-point of SE is unique.
\par
The SE framework of AMP works with any $P_X(x)$. Such $P_X(x)$ can be the distribution of, e.g., amplitude or phase modulation that is widely used signal transmission. For this reason, AMP is also suitable for communication applications such as massive MIMO detection \cite{Wu2014,Jeon2015}, and millimeter wave channel estimation \cite{Wen2015} (in which $\bm{A}$ represents a channel matrix). AMP has also been investigated for decoding sparse regression codes \cite{Rush2015,Barbier2015}, which have theoretically capacity approaching performances.
\par
The IID assumption for $\bm{A}$ is crucial to the SE of AMP \cite{Bayati2011,Bayati2015}. When $\bm{A}$ is not IID (especially when $\bm{A}$ is ill-conditioned), the accuracy of SE is not warranted and AMP may perform poorly \cite{Vila2014}. Various algorithms have been proposed to handle more general matrices \cite{Vila2014,Manoel2014,Rangan2015,Kabashima2014,Cakmak2014,Guo2015_UTAMP,Ccakmak2016}, but most of the existing algorithms lack accurate SE characterization. An exception is the work in \cite{Opper2016}, which considers a closely related problem and uses a method different from this paper. \par
The work in this paper is motivated by our observation that, the SE for AMP is still relatively reliable for a wider family of matrices other than IID Gaussian ones when the Onsager term is small. Our contributions are summarized below.
\begin{itemize}
\item We propose a modified AMP algorithm comprising of a de-correlated LE and a divergence-free NLE\footnote{The name is from \cite{Bostan2015}, although the discussions therein are irrelevant to this paper.}. The proposed algorithm allows LE structures beyond MF, such as pseudo-inverse (PINV) and linear MMSE (LMMSE). OAMP extends and provides new interpretations of our previous work in \cite{Yuan2014,Ma2015a}.
\item We derive an SE procedure for OAMP, which is accurate if the errors are independent during the iterative process. Independency, however, is a tricky condition. We will show that the use of a de-correlated LE and a divergence-free NLE makes the errors statistically orthogonal, hence the name orthogonal AMP (OAMP). Intuitively, such orthogonality partially satisfies the independency requirement. Our numerical results indicate that the SE predictions are reliable for various matrix ensembles (e.g., IID Gaussian, partial orthogonal and some ill-conditioned ones for which AMP does not work well) and also for various LE structures as mentioned above. Thus OAMP may have wider applications than AMP.
\item We derive optimal choices within the OAMP framework. We find that the fixed-point characterization of the SE is consistent with that of the optimal MMSE performance obtained by the replica method. This implies the potential optimality of OAMP. Compared with AMP, our result holds for the more general unitarily-invariant matrix ensemble.
\end{itemize}
We will provide numerical results to show that, compared with AMP, OAMP can achieve better MSE performance as well as faster convergence speed for ill-conditioned matrices. We will demonstrate the excellent performance of OAMP in communication systems with non-sparse binary phase shift keying (BPSK) signals as well as conventional sparse signals.\par
After we posted the preprint of this work \cite{Ma2016}, a proof was given for the state evolution of an OAMP related algorithm in systems involving unitarily-invariant matrices \cite{Rangan2016}.\par
Part of the results in this paper have been published in \cite{Ma2016ITW}. In this paper, we provide more detailed analysis and numerical results.
\par\vspace{5pt}
\textit{Notations:} Boldface lowercase letters represent vectors and boldface uppercase symbols denote matrices. $\mathbf{0}$ for a matrix or a vector with all-zero entries, $\mb{I}$ for the identity matrix with a proper size, $\bm{a}^{\mr{T}}$ for the conjugate of $\bm{a}$, $\|\bm{a}\|$ for the $\ell_2$-norm of the vector $\bm{a}$, $\ms{tr}(\bm{A})$ for the trace of $\bm{A}$, $\left( {\eta \left( {\bf{a}} \right)} \right)_j  \equiv \eta \left( {a_j } \right).$ $\ms{diag}\{\bm{A}\}$ for the diagonal part of $\bm{A}$, $\mathcal{N}(\bm{\mu},\bm{C})$ for Gaussian distribution with mean $\bm{\mu}$ and covariance $\bm{C}$, $\ms{E}\{\cdot\}$ for the expectation operation over all random variables involved in the brackets, except when otherwise specified. $\mr{E}\{a|b\}$ for the expectation of $a$ conditional on $b$, $\ms{var}\{{a}\}$ for $\ms{E}\left\{ \left({a} - \ms{E}\{{a}\}\right)^2 \right\}$, $\ms{var}\{{a}|{b}\}$ for $\ms{E}\left\{ \left({a} - \ms{E}\{{a}|{b}\}\right)^2|{b} \right\}$. 
\section{AMP}\label{Sec:II}
\subsection{AMP Algorithm}
Following the convention in \cite{Donoho2009}, assume that $\bm{A}$ is column normalized, i.e., $\mr{E}\{\|\bm{A}_{:,j}\|^2\approx1\}$ for each $j$. Approximate message passing (AMP) \cite{Donoho2009} refers to the following iterative process (initialized with $\bm{s}^0 = \bm{r}_{\mr{Onsager}}^0 = \mathbf{0}$)\footnote{The formulation here is different to the standard form in \cite{Donoho2009}, but they can be shown to be equivalent.}:
\BS \label{Eqn:II-B-AMP}
\begin{alignat}{3}
&\text{LE:}\qquad &&{\bf{r}}^t  = {\bf{s}}^t {\rm{ + }}{\bf{A}}^{\rm{T}} \left( {{\bf{y}}{\rm{ - }}{\bf{As}}^t } \right) + {\bf{r}}_{{\rm{Onsager}}}^t \label{Eqn:II-B-AMP-a} \\
&\text{NLE:}\qquad &&{\bf{s}}^{t + 1} = \eta _t \left( {{\bf{r}}^t } \right),
\end{alignat}
where $\eta_t$ is a component-wise Lipschitz continuous function of $\bm{r}^t$ and $\bm{r}_{\mr{Onsager}}^t$ an ``Onsager term" \cite{Donoho2009} defined by
\BE
\bm{r}_{\mr{Onsager}}^t  = \frac{N}{M} \cdot \bigg( \frac{1}{N}\sum_{j = 1}^N \eta_{t-1}'(r_{j}^{t-1}) \bigg) \cdot \left( {{\bf{r}}^{t - 1}  - {\bf{s}}^{t - 1} } \right).
\EE
\ES
The final estimate is $\bm{s}^{t+1}$.\par
The use of the Onsager term is the key to AMP. It regulates correlation during iterative processing and ensures the accuracy of SE when $\bm{A}$ has IID entries \cite{Donoho2009,Bayati2011}.
\subsection{State Evolution for AMP}
Define
\BS\label{Eqn:II-B-1}
\begin{equation}
{\bf{q}}^t  \equiv {\bf{s}}^t  - {\bf{x}}\text{ and } {\bf{h}}^t  \equiv {\bf{r}}^t  - {\bf{x}}.
\end{equation}
\ES
 After some manipulations, \eqref{Eqn:II-B-AMP} can be rewritten as \cite[Eqn.~(3.3)]{Bayati2011} (with initialization $\bm{q}^0=-\bm{x}$ and $\bm{h}^0_{\mr{Onsager}}=\mathbf{0}$):
 \BS\label{Eqn:II-B-AMP2}
\begin{alignat}{3}
&\text{LE:}\qquad&&{\bf{h}}^t  = \left( {{\bf{I}} - {\bf{A}}^{\rm{T}} {\bf{A}}} \right){\bf{q}}^t  + {\bf{A}}^{\rm{T}} {\bf{n}} + {\bf{h}}_{{\rm{Onsager}}}^t ,\\
&\text{NLE:}\qquad&&{\bf{q}}^{t + 1}  = \eta_t \left( \bm{x} +\bm{h}^t\right)-\bm{x}, \label{Eqn:II-B-AMP2-a}
\end{alignat}
where
\BE
{\bf{h}}_{{\rm{Onsager}}}^t  = \frac{N}{M} \cdot  \Bigg(\frac{1}{N} {\sum\limits_{j = 1}^N {\eta_{t - 1}' \left({x_j}+{h}_j^{t-1} \right)} } \Bigg) \cdot \left( {{\bf{h}}^{t - 1}  - {\bf{q}}^{t - 1} } \right),\label{Eqn:II-B-AMP2-c}
\EE
\ES
Strictly speaking, \eqref{Eqn:II-B-AMP2} is not an algorithm since it involves $\bm{x}$ that is to be estimated. Nevertheless, \eqref{Eqn:II-B-AMP2} is convenient for the analysis of AMP discussed below. \par
The SE for AMP refers to the following recursion:
\BS\label{Eqn:II-B-SE}
\begin{alignat}{3}
&\text{LE:}\qquad &&\tau _t^2  = \frac{N}{M} \cdot v_t^2 +\sigma ^2 , \label{Eqn:II-B-SE-a}\\
&\text{NLE:}\qquad && v_{t + 1}^2  = {\ms{E}}\left \{ \left [\eta_t \left( X+\tau _t Z \right)-X\right ]^2 \right \},
\end{alignat}
\ES
where $Z\sim\mathcal{N}(0,1)$ is independent of $X\sim P_X(x)$, and $v^2_0=\ms{E}\{X^2\}$. \par
When $\bm{A}$ has IID Gaussian entries, SE can accurately characterize AMP, as shown in Theorem 1 \cite{Bayati2011} below.
\begin{theorem}\cite[Theorem 2]{Bayati2011}
Let $\psi:\mathbb{R}^2\mapsto\mathbb{R}$ be a pseudo-Lipschitz function\footnote{The function $\psi$ is said to be pseudo-Lipschitz (or order two) \cite{Bayati2011} if there exists a constant $L>0$ such that for all $x$, $y$, $|\psi(x)-\psi(y)|\le L( 1 + \|x\| + \|y\| )\|x-y\|$.}. For each iteration, the following holds almost surely when $M,N\to\infty$ with a fixed ratio
\BE \label{Eqn:II-B-IIDGC}
\frac{1}{N}\sum\limits_{j = 1}^N {\psi \left( {h_j^t ,x_j } \right)}  \to {\ms{E}}\left\{ {\psi \left( {\tau _t Z},X \right)} \right\},
\EE
where $\tau_t$ is given in \eqref{Eqn:II-B-SE}.
\end{theorem}\par
To see the implication of Theorem 1,  let $\psi(h,x)\equiv\left[\eta_t(x+h)-x\right]^2$ in \eqref{Eqn:II-B-IIDGC}. Then, Theorem 1 says that the empirical mean square error (MSE) of AMP defined by
\BE \label{Eqn:AMP_MSE1}
\frac{1}{N}\left\| {\eta _t \left( {{\bf{x}} + {\bf{h}}^t } \right) - {\bf{x}}} \right\|^2
\EE
converges to the predicted MSE (where $\tau_t$ is obtained using SE) defined by
\BE \label{Eqn:AMP_MSE2}
{\ms{E}}\left\{ {\left[ {\eta _t \left( {X + \tau _t Z} \right) - X} \right]^2 } \right\}.
\EE
\subsection{Limitation of AMP}
The assumption that $\bm{A}$ contains IID entries is crucial to theorem 1. For other matrix ensembles, SE may become inaccurate. Here is an example. Consider the following function for the NLE in AMP\footnote{Strictly speaking, $\eta_t$ in \eqref{Eqn:II-D-eta} is not a component-wise function as required in AMP. However, if Theorem 1 holds, $\sum_{j=1}^N\hat{\eta}_t'(r_j^t)/N$ will converge to a constant independent of each individual $r_j^t$. In this case, $\eta_t$ is an approximate component-wise function and $\sum_{j=1}^N{\eta}_t'(r_j^t)/N\approx\beta\cdot\sum_{j=1}^N\hat{\eta}_t'(r_j^t)/N$.}
\BE \label{Eqn:II-D-eta}
\eta _t \left( {{\bf{r}}^t } \right) = \hat \eta _t \left( {{\bf{r}}^t } \right) - \left( {1 - \beta } \right) \cdot \bigg( {\frac{1}{N}\sum\limits_{j = 1}^N {\hat \eta _t' \left( {r_j^t } \right)} } \bigg) \cdot {\bf{r}}^t ,
\EE
where $\hat{\eta}_t$ is the thresholding function (which is commonly used in sparse signal recovery algorithms \cite{Donoho1995}) given in \eqref{Eqn:threshold} with $\gamma_t=1$. A family of  $\eta_t$ is obtained by changing $\beta$. In particular, $\eta_t$ reduces to the soft-thresholding function $\hat{\eta}_t$ when $\beta= 1$. We define a measure of the SE accuracy (after a sufficient number of iterations) as
\BE \label{Eqn:AMP_E}
E \equiv \frac{\left| {MSE}_{\mr{sim}} -{MSE}_{\mr{SE}}\right|}{{MSE}_{\mr{sim}}},
\EE
where $MSE_{\mr{sim}}$ and $MSE_{\mr{SE}}$ are the simulated and predicted MSEs in \eqref{Eqn:AMP_MSE1} and \eqref{Eqn:AMP_MSE2}. Here, as the empirical MSE is still random for large but finite $M$ and $N$, we average it over multiple realizations. \par
\begin{figure}[htbp]
\centering
  \includegraphics[width=.5\textwidth]{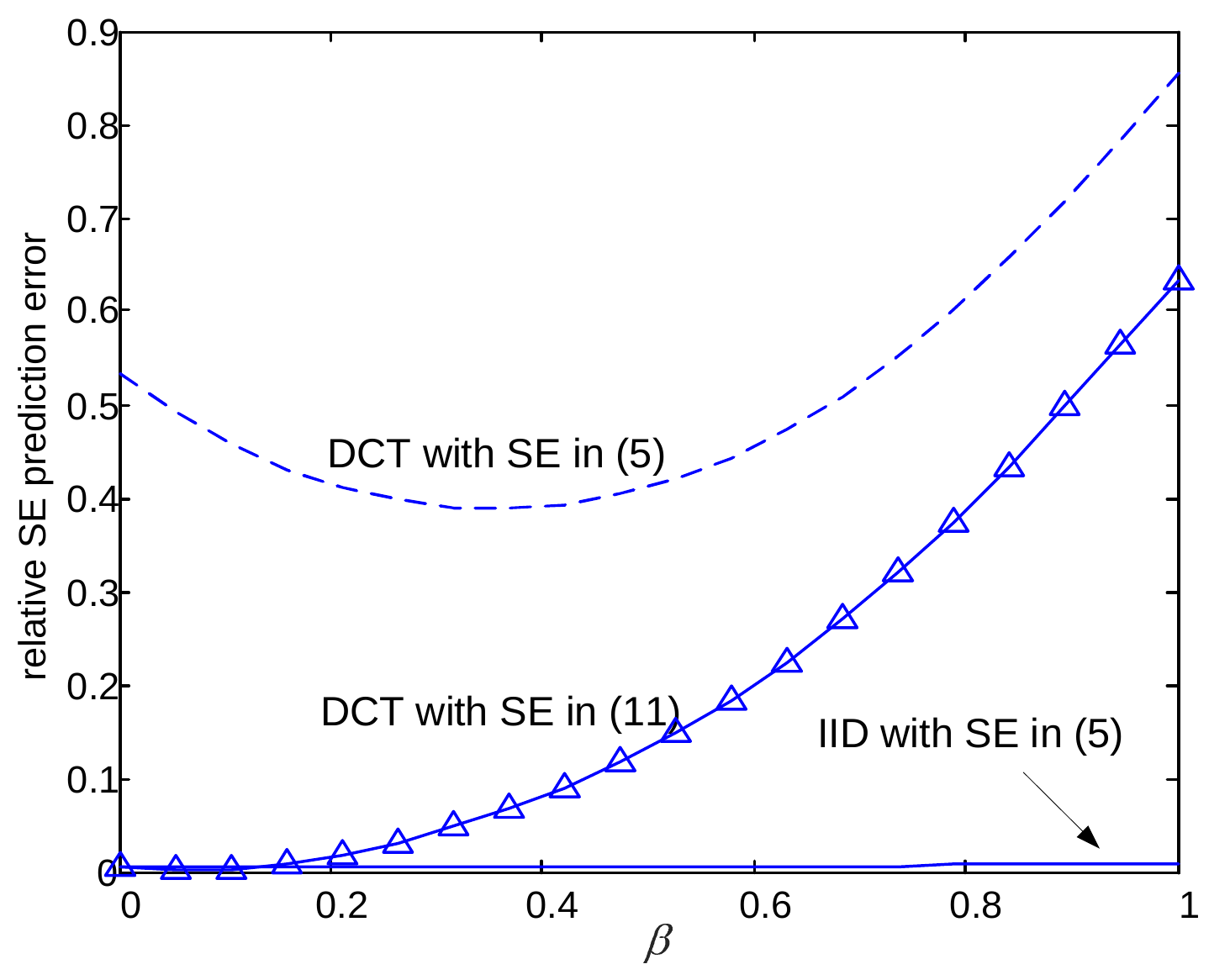}
  \caption{State evolution prediction error for AMP with a partial DCT matrix. $N = 8192$. $M = 5734(\approx0.7N)$. $SNR = 50$ dB. $\rho=0.4$. (See the signal model in Section~\ref{Sec:V}.) The simulated MSE is averaged over 100 independent realizations. The number of iterations is 50.}\label{Fig:SE-div}
\end{figure}
By changing $\beta$ from 0 to 1, we obtain a family of $\eta_t$. The solid line in Fig.~\ref{Fig:SE-div} shows $E$ defined in \eqref{Eqn:AMP_E} against $\beta$ for $\bm{A}$ being IID Gaussian. We can see that SE is quite accurate in the whole range of $\beta$ shown (with $E <10^{-2}$), which is consistent with the result in Theorem 1.\par
However, as shown by the dashed line, SE is not reliable when $\bm{A}$ is a partial DCT matrix. The partial DCT matrix can be obtained by uniformly randomly selecting the rows of a discrete cosine transform (DCT) matrix, and it is a widely used in compressed sensing. To see the problem, let us ignore the Onsager term. Suppose that $\bm{q}^t$ consists of IID entries with $\mr{E}\big\{(q_j^t)^2\big\}=v_t^2$, and $\bm{q}^t$ is independent of $\bm{A}$ and $\bm{n}$. It can be verified that
\BE \label{Eqn:AMP_SE_new}
\tau_t^2\equiv\frac{1}{N}\mr{E}\left\{\|\bm{h}^t\|^2\right\}=\frac{N-M}{M}\cdot v_t^2+\sigma^2.
\EE
Clearly, this is inconsistent with the SE in \eqref{Eqn:II-B-SE-a}. The problem is caused by the discrepancy in eigenvalue distributions: \eqref{Eqn:AMP_SE_new} above is derived from the eigenvalue distribution of a partial DCT matrix while \eqref{Eqn:II-B-SE-a} from that of an IID Gaussian $\bm{A}$.\par
How about replacing \eqref{Eqn:II-B-SE-a} by \eqref{Eqn:AMP_SE_new} for the partial DCT matrix? This is shown by the solid line with triangle markers in Fig.~\ref{Fig:SE-div}. We can see that $E$ is still large for $ \beta> 0$, which can be explained by the fact the Onsager term was ignored above. Interestingly, we can see that $E$ is very small at $\beta=0$, where the Onsager term vanishes for the related $\eta_t$ in \eqref{Eqn:II-D-eta}. This observation motivates the work presented below.

\section{Orthogonal AMP}\label{Sec:III}
In this section, we first introduce the concepts for de-correlated and divergence-free structures for the LE and NLE. We then discuss the OAMP algorithm and its properties.
\subsection{De-correlated Linear Estimator}\label{Sec:III-A}
Return to \eqref{Eqn:Intro_a}: $\bm{y}=\bm{Ax}+\bm{n}$. Let $\bm{s}$ be an estimate of $\bm{x}$. Assume that $\bm{s}$ has IID entries with $\mr{E}\{(s_j-x_j)^2\}=v^2$. Consider the linear estimation (LE) structure below \cite{Kay1993} for $\bm{x}$
\BE
\bm{r} = \bm{s} + \bm{W}(\bm{y}-\bm{As}),
\EE
which is specified by $\bm{W}$. Let the singular value decomposition (SVD) of $\bm{A}$ be $\bm{A=V\Sigma U}^{\mr{T}}$. Throughput this paper, we will focus on the following structure for $\bm{W}$
\BE \label{Eqn:OAMP_W}
\bm{W}=\bm{UGV}^{\mr{T}}.
\EE
\begin{definition}[Unitarily-invariant matrix]
$\bm{A=V\Sigma U}^{\mr{T}}$ is said unitarily-invarint \cite{Tulino2004} if $\bm{U}$, $\bm{V}$ and $\bm{\Sigma}$ are mutually independent, and $\bm{U}$, $\bm{V}$ are Haar-distributed (i.e., isotropically random orthogonal).\footnote{It turns out that the distribution of $\bm{V}$ does not affect the average performance of OAMP. The reason is that OAMP implicitly estimates $\bm{x}$ based on $\bm{V}^{\mr{T}}\bm{y}$, and $\bm{V}^{\mr{T}}\bm{n}$ has the same distribution as $\bm{n}$ for an arbitrary orthogonal matrix $\bm{V}$ due to the unitary-invariance of Gaussian distribution \cite{Tulino2004}.}
\end{definition}\par \vspace{5pt}
Assume that $\bm{A}$ is unitarily-invariant. We will say that the LE (or $\bm{W}$ in \eqref{Eqn:OAMP_W}) is a de-correlated one if $\mr{tr}(\bm{I}-\bm{WA})=0$. Given an arbitrary $\hat{\bm{W}}$ that satisfies \eqref{Eqn:OAMP_W}, we can construct $\bm{W}$ with $\mr{tr}(\bm{I}-\bm{WA})=0$ as follows
\BE \label{Eqn:OAMP_W_decorr}
\bm{W}=\frac{N}{\mr{tr}(\hat{\bm{W}}\bm{A})}\bm{\hat{\bm{W}}}.
\EE
The following are some common examples \cite{Kay1993} of such $\hat{\bm{W}}$\vspace{5pt}
\BS \label{Eqn:OAMP_LEs}
matched filter (MF):
\BE  \label{Eqn:OAMP_MF}
\hat{\bm{W}}^{\mr{MF}}  =\bm{A}^{\mr{T}},
\EE
pseudo-inverse (PINV)\footnote{We assume that $\bm{A}$ has full rank.}:
\BE \label{Eqn:OAMP_PINV}
\hat{\bm{W}}^{\mr{PINV}}
=
\begin{cases}
\bm{A}^{\mr{T}}(\bm{AA}^{\mr{T}})^{-1} &\text{if } M<N\\
\left(\bm{A}^{\mr{T}}\bm{A}\right)^{-1}\bm{A}^{\bm{T}} & \text{if }M>N,
\end{cases}
\EE
linear MMSE (LMMSE):
\BE \label{Eqn:OAMP_LMMSE}
\hat{\bm{W}}^{\mr{LMMSE}}=v^2\bm{A}^{\mr{T}}(v^2\bm{AA}^{\mr{T}}+\sigma^2\mb{I})^{-1}. 
\EE
\ES\par
We will discuss the properties of de-correlated LE in Section \ref{Sec:III-F} later.
\subsection{Divergence-free Estimator} \label{Eqn:III-A}
Consider signal estimation from an observation corrupted by additive Gaussian noise
\BE \label{Eqn:III-A-AWGN}
R = X + \tau Z,
\EE
where $X\sim P_X(x)$ is the signal to be estimated and is independent of $Z\sim\mathcal{N}(0,1)$. For this additive Gaussian noise model, we define divergence-free estimator (or a divergence-free function of $R$) as follows.
\begin{definition}[Divergence-free Estimator]
We say $\eta:\mathbb{R}\mapsto\mathbb{R}$ is divergence-free (DF) if
\BE
{\ms{E}}\left\{ {\eta {\rm{'}}\left( R \right)} \right\} = 0.
\EE
\end{definition} \par \vspace{5pt}
A divergence-free function $\eta$ can be constructed as
\BE \label{Eqn:III-A-DF-fun}
\eta \left( r \right) = C \cdot \left( {\hat \eta \left( r \right) - \mathop {\rm{E}}\limits_R \left\{ {\hat \eta {\rm{'}}\left( R \right)} \right\} \cdot r} \right),
\EE
where $\hat{\eta}$ is an arbitrary function and $C$ an arbitrary constant.
\subsection{OAMP Algorithm}\label{Sec:III-C}
Starting with $\bm{s}^0 = \mathbf{0}$, OAMP proceeds as
\BS \label{Eqn:III-B-OAMP}
\begin{alignat}{3}
&\text{LE:}\qquad &&{\bf{r}}^t  = {\bf{s}}^t  + {\bf{W}}_t \left( {{\bf{y}} - {\bf{As}}^t } \right), \label{Eqn:III-B-OAMP-a} \\
&\text{NLE:}\qquad &&{\bf{s}}^{t + 1}  = \eta _t \left( {{\bf{r}}^t } \right),
\end{alignat}
\ES
where $\bm{W}_t$ is de-correlated and $\eta_t$ is divergence-free. In the final stage, the output is
\BE
\left({\bf{s}}^{t + 1}\right)^{\mr{out}}  = \eta _t^{{\rm{out}}} \left( {{\bf{r}}^t } \right),
\EE
where $\eta _t^{{\rm{out}}}$ is not necessarily divergence-free.\par
OAMP is different from the standard AMP in the following aspects:
\begin{itemize}
\item In \eqref{Eqn:III-B-OAMP-a}, $\bm{W}_t$ is restricted to be de-correlated,  but it still has more choices than its counterpart $\bm{A}^{\mr{T}}$ in \eqref{Eqn:II-B-AMP-a}\footnote{When the entries of $\bm{A}$ are IID with zero mean and variance $1/M$ (as considered in \cite{Donoho2009}), $N/\mr{tr}(\bm{A}^{\mr{T}}\bm{A})\approx1$, and so $\bm{W}_t=\mb{A}^{\mr{T}}$ satisfies the condition in \eqref{Eqn:OAMP_W} and \eqref{Eqn:OAMP_W_decorr}.}.
\item In \eqref{Eqn:III-B-OAMP-a},  the function $\eta_t$ is restricted to be divergence-free. Consequently, the Onsager term vanishes.
\item A different estimation function $\eta_t^{\mr{out}}$ (not necessarily divergence-free) is used to produce a final estimate.
\end{itemize}
We will show that, under certain assumptions, restricting $\bm{W}_t$ to be de-correlated and $\eta_t$ to be divergence-fee ensure the orthogonality between the input and output ``error'' terms for both LE and NLE. The name 	``orthogonal AMP'' comes from this fact.
\subsection{OAMP Error Recursion and SE}\label{Sec:III-D}
Similar to \eqref{Eqn:II-B-1}, define the error terms as $\bm{h}^t\equiv\bm{r}^t-\bm{x}$ and $\bm{q}^t\equiv\bm{s}^t-\bm{x}$. We can write an error recursion for OAMP (similar to that for AMP in \eqref{Eqn:II-B-AMP2}) as
\BS \label{Eqn:OAMP_error}
\begin{alignat}{3}
&\text{LE:}\qquad &&\bm{h}^t=\bm{B}_t\bm{q}^t+\bm{W}_t\bm{n} \label{Eqn:OAMP_error_a} \\
&\text{NLE:}\qquad && \bm{q}^{t+1}=\eta_t(\bm{x}+\bm{h}^t)-\bm{x}, \label{Eqn:OAMP_error_b}
\end{alignat}
\ES
where $\bm{B}_t\equiv\bm{I}-\bm{W}_t\bm{A}$. Two error measures are introduced:
\BS
\begin{align}
\tau_t^2&\equiv\frac{1}{N}\cdot\mr{E}\left\{\|\bm{h}^t\|^2\right\},\\
v_{t+1}^2&\equiv\frac{1}{N}\cdot\mr{E}\left\{\|\bm{q}^{t+1}\|^2\right\}.
\end{align}
\ES
The SE for OAMP is defined by the following recursion
\BS \label{Eqn:OAMP_SE}
\begin{alignat}{3}
&\text{LE:}\  &&\tau_t^t=\frac{1}{N}\mr{E}\left\{\mr{tr}(\bm{B}_t\bm{B}_t^{\mr{T}})\right\} v_t^2
+\frac{1}{N}\mr{E}\left\{\mr{tr}(\bm{W}_t\bm{W}_t^{\mr{T}})\right\} \sigma^2 \label{Eqn:OAMP_SE_a}\\
& \text{NLE:}\  && v_{t+1}^2=\mr{E}\left\{\left[\eta_t(X+\tau_t Z)-X\right]^2\right\}, \label{Eqn:OAMP_SE_b}
\end{alignat}
\ES
where $X\sim P_X(x)$ is independent of $Z\sim\mathcal{N}(0,1)$. Also, at the final stage, the MSE is predicted as
\BE \label{Eqn:OAMP_SE_c}
\mr{E}\left\{\left[\eta_t^{\mr{out}}(X+\tau_t Z)-X\right]^2\right\}.
\EE
\subsection{Rationales for OAMP}\label{Sec:III-E}
It is straightforward to verify that the SE in \eqref{Eqn:OAMP_SE} is consistent with the error recursion in \eqref{Eqn:OAMP_error}, provided that the following two assumptions hold for every $t$.\vspace{5pt}
\begin{assumption}
$\bm{h}^t$ in \eqref{Eqn:OAMP_error_a} consists of IID zero-mean Gaussian entries independent of $\bm{x}$.
\end{assumption}
\begin{assumption}
$\bm{q}^{t+1}$ in \eqref{Eqn:OAMP_error_b} consists of IID entries independent of $\bm{A}$ and $\bm{n}$.
\end{assumption}\vspace{5pt}
According to our earlier assumption below \eqref{Eqn:Intro}, $\bm{x}$ is IID and independent of $\bm{A}$ and $\bm{n}$. In OAMP, $\bm{q}^{0}=-\bm{x}$, so Assumption 2 holds for $t=-1$. Thus the two Assumptions will hold if we can prove that they imply each other in the iterative process. Unfortunately, so far, we cannot.\par
Assumptions 1 and 2 are only sufficient conditions for the SE. Even if they do not hold exactly, the SE may still be valid. In Section \ref{Sec:V}, we will show that the SE for OAMP is accurate for a wide range of sensing matrices using simulation results.  In the following two subsections, we will see that, with a de-correlated $\bm{W}_t$ and a divergence-free $\eta_t$, Assumptions 1 and 2 can partially imply each other. We emphasize that the discussions below are to provide intuitions for OAMP, which are by no means rigorous.
\subsection{Intuitions for the LE Structure}\label{Sec:III-F}
Eqn.~\eqref{Eqn:III-B-OAMP-a} performs linear estimation of $\bm{x}$ from $\bm{y}$ based on Assumption 2 (for $\bm{q}^t$). We first consider ensuring Assumption 1 based on Assumption 2. The independence requirements in Assumption 1 are difficult to handle. We reduce our goal to remove the correlation among the variables involved. This is achieved by restricting $\bm{W}_t$ to be de-correlated, as shown below.
\begin{proposition}\label{Pro:IV}
 Suppose that Assumption 2 holds and $\bm{A}$ is unitarily-invariant. If $\bm{W}_t$ is de-correlated, then the entries of $\bm{h}^t$ are uncorrelated with those of $\bm{x}$. Furthermore, the entries of $\bm{h}^t$ in \eqref{Eqn:OAMP_error_a} are mutually uncorrelated with zero-mean and identical variances.
\end{proposition}
\begin{IEEEproof}
See Appendix \ref{App:I}.
\end{IEEEproof}\par
Some comments are in order.
\begin{enumerate}
\item [(i)]The name ``de-correlated'' LE comes from Proposition~\ref{Pro:IV}.
\item [(ii)] Under the same conditions as Proposition 1, the input and output error vectors for LE are uncorrelated, namely, $\mr{E}\left\{\bm{h}^t\left(\bm{q}^t\right)^{\mr{T}}\right\} = \mathbf{0}$.
\item [(iii)] A key condition to Proposition 1 is that the sensing matrix $\bm{A}$ is unitarily invariant. Examples of such $\bm{A}$ include the IID Gaussian matrix ensemble and the partial orthogonal ensemble \cite{Tulino2013}. Note that there is no restriction on the eigenvalues of $\bm{A}$. Thus, OAMP is potentially applicable to a wider range of $\bm{A}$ than AMP.
\item [(iv)] We can meet the de-correlated constraint using \eqref{Eqn:OAMP_W_decorr}, in which $\hat{\bm{W}}_t$ can be chosen from those in \eqref{Eqn:OAMP_LEs}. Thus OAMP has more choices for the LE than AMP, which makes the former potentially more efficient.
\end{enumerate}
\subsection{Intuitions for the NLE Structure}\label{Sec:III-G}
We next consider ensuring Assumption 2 based on Assumption 1. From \eqref{Eqn:OAMP_error}, if $\bm{q}^{t+1}$ is independent of $\bm{h}^t$, then it is also independent of $\bm{A}$ and $\bm{n}$, which can be seen from the Markov chain $\bm{A},\bm{n}\to\bm{h}^t\to\bm{q}^{t+1}$. Thus it is sufficient to ensure the independency between $\bm{q}^{t+1}$ and $\bm{h}^t$. Similar to the discussion in Section~\ref{Sec:III-F}, we reduce our goal to ensuring orthogonality between $\bm{q}^{t+1}$ and $\bm{h}^t$.\par
Suppose that Assumption 1 holds, we can construct an approximate divergence-free function $\eta_t$ according to \eqref{Eqn:III-A-DF-fun}:
\BE \label{Eqn:III-B-OAMP-DF}
\eta _t \left( {{\bf{r}}^t } \right) =C_t  \cdot \Bigg( {\hat \eta _t\left( {{\bf{r}}^t } \right) - \bigg( {\frac{1}{N}\sum\limits_{j = 1}^N {\hat \eta _t' \big( {r_j^t } \big)} } \bigg) \cdot {\bf{r}}^t } \Bigg).
\EE
All the numerical results about OAMP shown in Section \ref{Sec:V} are based on \eqref{Eqn:III-B-OAMP} and \eqref{Eqn:III-B-OAMP-DF}.\par
There is an inherent orthogonality property associated with divergence-free functions.
\begin{proposition}
If $\eta$ is a divergence-free function, then
\BE \label{Eqn:III-A-orth2}
{\rm{E}}\left\{ {\tau_t Z \cdot \eta \left( X+\tau_t Z \right)} \right\} = 0.
\EE
\end{proposition}
\begin{IEEEproof}
From Stein's Lemma \cite{Stein1972,Bayati2011}, we have
\BE \label{Eqn:III-A-Stein}
{\rm{E}}\left\{ {Z \cdot \varphi \left( Z \right)} \right\} = {\rm{E}}\left\{ {\varphi '\left( Z \right)} \right\},
\EE
for any $\varphi:\mathbb{R}\mapsto\mathbb{R}$ such that the expectations in \eqref{Eqn:III-A-Stein} exist. Applying Stein's lemma in \eqref{Eqn:III-A-Stein} with $\psi(Z)\equiv\eta_t(X+\tau_t Z)$, we have
\BS \label{Eqn:III-A-orth1}
\begin{align}
&{\rm{E}}\left\{ {\tau_t Z \cdot \eta_t \left(X+\tau_t Z \right)} \right\} \\
&= \tau_t  \cdot \mathop {\rm{E}}\limits_X \bigg\{ {\mathop {\rm{E}}\limits_{Z|X} \left\{ {Z \cdot \eta_t \left( {X + \tau_t Z} \right)} \right\}} \bigg\} \\
& = \tau_t ^2  \cdot \mathop {\rm{E}}\limits_X \bigg\{ {\mathop {\rm{E}}\limits_{Z|X} \left\{ {\eta_t '\left( {X + \tau_t Z} \right)} \right\}} \bigg\}  \label{Eqn:III-A-orth1-b}\\
& = \tau_t ^2  \cdot {\rm{E}}\left\{ {\eta_t '\left( {X + \tau_t Z} \right)} \right\},
\end{align}
\ES
where $\eta_t'(X+\tau_t Z)\equiv \eta_t'(R)|_{R=X+\tau_t Z}$. Combining \eqref{Eqn:III-A-orth1} with Definition 2, we arrive at \eqref{Eqn:III-A-orth2}.
\end{IEEEproof}\par\vspace{5pt}
Noting that $\mr{E}\{ZX\} = 0$, \eqref{Eqn:III-A-orth2} is equivalent to
\BE \label{Eqn:III-A-orth3}
{\rm{E}}\left\{ {\left( {R^t - X} \right) \cdot \left[ {\eta_t \left( R^t \right) - X} \right]} \right\} = 0,
\EE
where $R^t\equiv X+\tau_t Z$. In \eqref{Eqn:III-A-orth3}, $R^t - X$ and $\eta_t(R^t)-X$ represent, respectively, the error terms before and after the estimation. Eqn.~\eqref{Eqn:III-A-orth3} indicates that these two error terms are orthogonal. (They are also uncorrelated as $R^t-X$ has zero mean.) Thus the divergence-free constrain on the NLE is to establish orthogonality between $\bm{q}^{t+1}$ and $\bm{h}^t$.
\subsection{Brief Summary}
If the input and output errors of the LE and NLE are independent of each other, Assumptions 1 and 2 naturally hold. However, independency is generally a tricky issue. We thus turn to orthogonality instead. The name ``orthogonal AMP'' came from this fact. Propositions 1 and 2 are weaker than Assumptions 1 and 2. Nevertheless, our extensive numerical study (see Section~\ref{Sec:V}) indicates that the SE in \eqref{Eqn:OAMP_SE} is indeed reliable for OAMP. \par
Also note that each of Propositions 1 and 2 depends on one assumption, so they do not ensure orthogonality in the overall process. Nevertheless, we observed from numerical results that the orthogonality property is accurate for with unitarily-invariant matrices.
\subsection{MSE Estimation}
The MSEs $v_t^2\equiv\mr{E}\{\|\bm{q}^t\|^2\}/N$ and $\tau_t^2\equiv\mr{E}\{\|\bm{h}^t\|^2\}/N$ can be used as parameters of $\bm{W}_t$ and $\eta_t$. An example is the optimized $\bm{W}_t$ and $\eta_t$ given in Lemma 1 in Section~\ref{Sec:IV}. We now discuss empirical estimators for $v_t^2$ and $\tau_t^2$.\par
We can adopt the following estimator \cite[Eqn.~(71)]{Vila2013} for $v_t^2$
\BE \label{Eqn:IV-C-MSE1}
\hat v_t^2  = \frac{{\left\| {{\bf{y}} - {\bf{As}}^t } \right\|{}^2 - M \cdot \sigma ^2 }}{{{\rm{tr}}\left( {{\bf{A}}^{\rm{T}} {\bf{A}}} \right)}}.
\EE
Note that $\hat{v}_t^2$ in \eqref{Eqn:IV-C-MSE1} can be negative. We may use $\max(\hat{v}_t^2,\epsilon)$ as a practical estimator for $v_t^2$, where $\epsilon$ is a small positive constant. (Setting $\epsilon=0$ may cause a stability problem.)\par
Given $\hat{v}_t^2$, $\tau_t^2$ can be estimated using \eqref{Eqn:OAMP_SE_a}:
\BE \label{Eqn:IV-C-MSE2}
\hat{\tau}_t^t=\frac{1}{N}\mr{tr}(\bm{B}_t\bm{B}_t^{\mr{T}})\cdot \hat{v}_t^2
+\frac{1}{N}\mr{tr}(\bm{W}_t\bm{W}_t^{\mr{T}})\cdot \sigma^2.
\EE
In certain cases, Eqn.~\eqref{Eqn:IV-C-MSE2} can be simplified to more concise formulas. For example, \eqref{Eqn:IV-C-MSE2} simplifies to $\hat \tau _t^2  = \left( {N - M} \right)/M \cdot \hat v_t^2  + N/M^2  \cdot {\rm{tr}}\big\{ {\left( {{\bf{AA}}^{\rm{T}} } \right)^{ - 1} } \big\} \cdot \sigma ^2$ when $\bm{W}_t$ is given by the PINV estimator in \eqref{Eqn:OAMP_PINV} together with \eqref{Eqn:OAMP_W_decorr}. Also, simple closed-form asymptotic expression exists for \eqref{Eqn:IV-C-MSE2} for certain matrix ensembles. For example, \eqref{Eqn:OAMP_SE_a} converges to \eqref{Eqn:Numerical_SE_a},  \eqref{Eqn:Numerical_SE_b} and \eqref{Eqn:Numerical_SE_c} for IID Gaussian matrices with MF, PINV and LMMSE linear estimators, respectively.\par
The numerical results presented in Section~\ref{Sec:V} are obtained based on approximations in \eqref{Eqn:IV-C-MSE1} and \eqref{Eqn:IV-C-MSE2}.

\section{Optimization Structures for OAMP}\label{Sec:IV}
In this section, we derive the optimal LE and NLE structures for OAMP based on SE. We show that OAMP can potentially achieve optimal performance, provided that its SE is reliable.
\subsection{Asymptotic Expression for SE}
Recall that $\bm{A}=\bm{V\Sigma U}^{\mr{T}}$ and $\bm{B}=\bm{I-W}_t\bm{A}$. From \eqref{Eqn:OAMP_W} and \eqref{Eqn:OAMP_W_decorr}, we have $\bm{W}_t=N/\mr{tr}(\hat{\bm{W}}_t\bm{A})\cdot\hat{\bm{W}}_t$ and $\hat{\bm{W}}_t=\bm{U}\hat{\bm{G}}_t\bm{V}^{\mr{T}}$. With these definitions, we can rewrite the right hand side of \eqref{Eqn:OAMP_SE_a} as follows
\BE \label{Eqn:opt_SE_exp}
\Phi_t(v_t^2)\equiv\left(      \frac{ \frac{1}{N} \sum_{i=1}^N \hat{g}_i^2\lambda_i^2}{\left(\frac{1}{N} \sum_{i=1}^N \hat{g}_i\lambda_i\right)^2  }  -1 \right) v_t^2
+\left(   \frac{ \frac{1}{N} \sum_{i=1}^N \hat{g}_i^2}{\left(\frac{1}{N} \sum_{i=1}^N \hat{g}_i\lambda_i\right)^2  }\right) \sigma^2,
\EE
where $\lambda_i$ and $\hat{g}_i$ ($i=1,\ldots,M$) denote the $i$th diagonal entries of $\bm{\Sigma}$ ($M\times N$) and $\hat{\bm{G}}_t$ ($N\times M$), respectively. In \eqref{Eqn:opt_SE_exp}, we define $\lambda_i=\hat{g}_i=0$ for $i=M+1,\ldots,N$). \par
In \eqref{Eqn:opt_SE_exp}, $\Phi_t(v_t^2)$ is for fixed $\{\lambda_i\}$ and $\{\hat{g}_i\}$. Now, following \cite{Vehkapera2014}, assume that the empirical cumulative distribution function (cdf) of $\{\lambda_1^2,\ldots,\lambda_N^2\}$, denoted by
\BE \label{Eqn:opt_CDF}
\hat{F}_{\bm{A}^{\mr{T}}\bm{A}}(\lambda^2)=\frac{1}{N}\sum_{i=1}^N\mathbb{I}(\lambda_i^2\ge \lambda^2)
\EE
converges to a limiting distribution when $M,N\to\infty$ with a fixed ratio. Furthermore, assume that $\hat{g}_i$ can be generated from $\lambda_i$ as $\hat{g}_i=\hat{g}_t(v_t^2,\lambda_i)$ with $\hat{g}_t$ a real-valued function. Then, \eqref{Eqn:opt_SE_exp} converges to
\BE \label{Eqn:opt_SE_asy}
\Phi_t(v_t^2)\to \left(  \frac{\mr{E}\{\hat{g}_t^2\lambda^2\}}{(\mr{E}\{\hat{g}_t\lambda\})^2} -1\right)\cdot v_t^2
+  \frac{\mr{E}\{\hat{g}_t^2\}}{(\mr{E}\{\hat{g}_t\lambda\})^2}\cdot \sigma^2,
\EE
where the expectations (assumed to exist) are taken over the asymptotic eigenvalue distribution of $\bm{A}^{\mr{T}}\bm{A}$ (\textit{including the zero eigenvalues}) and $\hat{g}_t$ stands for $\hat{g}_t(v_t^2,\lambda)$.\par
We further define
\BE \label{Eqn:opt_SE_psi}
\Psi_t(\tau_t^2)\equiv\mr{E}\left\{ \left[\eta_t(X+\tau_t Z)-X\right]^2 \right\},
\EE
where $\eta_t(r)\equiv C_t\cdot\left[ \hat{\eta}_t(r)-\mr{E}\{\hat{\eta}_t'(X+\tau_t Z)\}\cdot r \right]$ and $X$ is independent of $Z\sim\mathcal{N}(0,1)$. Then, from \eqref{Eqn:opt_SE_exp}, \eqref{Eqn:OAMP_SE_b} and \eqref{Eqn:opt_SE_psi}, the SE for OAMP is given by (with $v_0^2=\mr{E}\{X^2\}$)
\BS  \label{Eqn:opt_SE}
\begin{alignat}{3}
&\text{LE:}\qquad &&\tau_t^2 &= \Phi_t(v_t^2) \label{Eqn:opt_SE_a},\\
&\text{NLE:}\qquad && v_{t+1}^2&=\Psi_t(\tau_t^2). \label{Eqn:opt_SE_b}
\end{alignat}
\ES \par
The estimate for $\bm{x}$ in OAMP is generated by $\eta_t^{\mr{out}}$ rather than $\eta_t$. Thus, the MSE performance of OAMP, measured by $\|\eta_t^{\mr{out}}(\bm{r}^t)-\bm{x}\|^2/N$, is predicted as
\BE \label{Eqn:opt_SE_psi_out}
\Psi_t^{\mr{out}}(\tau_t^2)\equiv\mr{E}\left\{  \left[ \eta_t^{\mr{out}}(X+\tau_t Z)-X\right]^2 \right\}.
\EE
\subsection{Optimal Structure of OAMP}
We now derive the optimal $\bm{W}_t$, $\eta_t$ and $\eta_t^{\mr{out}}$ that minimize the MSE at the final iteration.\par
Let $\Phi_t^{\star}$, $\Psi_t^{\star}$, and $(\Psi_t^{\mr{out}})^{\star}$ be the minimums of $\Phi_t$, $\Psi_t$, and $\Psi_t^{\mr{out}}$ respectively (the minimizations are taken over $\bm{W}_t$, $\eta_t$, and $\eta_t^{\mr{out}}$). Lemmas 1 and 2 below will be useful to prove Theorem 2.
\begin{lemma}
The optimal $\bm{W}_t$ and $\eta_t$ that minimize $\Phi_t$ and $\Psi_t$ in \eqref{Eqn:opt_SE_exp} and \eqref{Eqn:opt_SE_psi} are given by
\BS \label{Eqn:V-A-opt}
\BE \label{Eqn:V-A-opt-a}
\bm{W}_t^{\star}= \frac{N}{\mr{tr}(\hat{\bm{W}}_t^{\mr{LMMSE}}\bm{A})}\hat{\bm{W}}_t^{\mr{LMMSE}},
\EE
\BE  \label{Eqn:V-A-opt-c}
\eta _t^{\star}(R^t) = C_t^{\star}  \cdot \left( { {\eta _t^{{\rm{MMSE}}} } (R^t) - \frac{{mmse_B \left( {\tau _t^2 } \right)}}{{\tau _t^2 }} \cdot R^t } \right),
\EE
where
\BE \label{Eqn:V-A-opt-d}
C_t^{\star}  \equiv \frac{{\tau _t^2 }}{{\tau_t^2  - mmse_B \left( {\tau _t^2 } \right)}},
\EE
\BE \label{Eqn:V-A-opt-b}
{\eta _t^{{\rm{MMSE}}} }(R^t)  = {\rm{E}}\left\{ {X|R^t  = X + \tau _t Z} \right\},
\EE
\BE \label{Eqn:V-A-opt-e}
mmse_B \left( {\tau _t^2 } \right) \equiv {\rm{E}}\left\{ \left(\eta_t^{\mr{MMSE}}-X\right)^2 \right\}.
\EE
\ES
Furthermore, the optimal $(\eta_t^{\mr{out}})^{\star}$ that minimizes $\Psi_t^{\mr{out}}$ is given by $\eta_t^{\mr{MMSE}}$.
\end{lemma} \vspace{5pt}
\begin{IEEEproof}
The optimality of $(\eta_t^{\mr{out}})^{\star}$ is by definition. The optimality of $\bm{W}_t^{\star}$ and $\eta_t^{\star}$ are not so straightforward, due to the de-correlated constraint on $\bm{W}_t$ and the divergence-free constraint on $\eta_t$. The details are given in Appendix~\ref{App:II}.
\end{IEEEproof}\par
Substituting $\bm{W}_t^{\star}$, $\eta_t^{\star}$ and $(\eta_t^{\mr{out}})^{\star}$ into \eqref{Eqn:opt_SE_exp}, \eqref{Eqn:opt_SE_psi} and \eqref{Eqn:opt_SE_psi_out}, and after some manipulations, we obtain
\BS \label{Eqn:opt_SE_opt}
\begin{alignat}{3}
&\text{LE:}\qquad &&\Phi^{\star}(v_t^2)  = \left( \frac{1}{mmse_A(v_t^2)} - \frac{1}{v_t^2}\right)^{-1}, \label{Eqn:opt_SE_opt_a}\\
&\text{NLE:}\qquad &&\Psi^{\star}(\tau_t^2)  = \left( \frac{1}{mmse_B(\tau_t^2)} - \frac{1}{\tau_t^2}\right)^{-1},  \label{Eqn:opt_SE_opt_b}\\
&\text{NLE:}\qquad&&\left(  \Psi^{\mr{out}} \right)^{\star}(\tau_t^2)  = mmse_B(\tau_t^2), \label{Eqn:opt_SE_opt_c}
\end{alignat}
\ES
where $mmse_A(v_t^2)\equiv\frac{1}{N}\sum_{i=1}^N \frac{\sigma^2\cdot v_t^2}{v_t^2\cdot\lambda_i^2+\sigma^2}$ and $mmse_B(\tau_t^2)$ is given in \eqref{Eqn:V-A-opt-e}. The derivations of \eqref{Eqn:opt_SE_opt_a} are omitted, and the derivations for \eqref{Eqn:opt_SE_opt_b} are shown in Appendix~\ref{App:III-A}. In \eqref{Eqn:opt_SE_opt}, the subscript $t$ has been omitted for
the functions $\Phi^{\star}$, $\Psi^{\star}$ and $(\Psi^{\mr{out}})^{\star}$ as they do not change across iterations. \par
\begin{lemma}
The functions $\Phi^{\star}$, $\Psi^{\star}$, and $(\Psi^{\mr{out}})^{\star}$ in \eqref{Eqn:opt_SE_opt} are monotonically increasing.
\end{lemma}\par\vspace{5pt}
\begin{IEEEproof}
The monotonicity of $(\Psi^{\mr{out}})^{\star}$ follows directly from the monotonicity of MMSE for additive Gaussian noise models \cite{Guo2011}. The monotonicity of $\Phi^{\star}$ and $\Psi^{\star}$ are proved in Appendix~\ref{App:III-B}.
\end{IEEEproof}
According to the state evolution process, the final MSE can be expressed as
\BE \label{Eqn:V-A-2}
\Psi _t^{{\rm{out}}} \left( {\Phi _t \left( {\Psi _{t - 1} \left( {\Phi _{t - 1} \left( { \cdots \left( {\Phi _0 \left( {v_0^{\rm{2}} } \right)} \right) \cdots } \right)} \right)} \right)} \right).
\EE
From Lemmas 1 and 2, replacing any function (i.e., $\{\Phi_{t'}\}$, $\{\Psi_{t'}\}$, and $\Psi_{t}^{\mr{out}}$) in \eqref{Eqn:V-A-2} by its local minimum reduces the final MSE. This leads to the following theorem.
\begin{theorem}
For the SE in \eqref{Eqn:opt_SE}, the final MSE in \eqref{Eqn:V-A-2} is minimized by $\{\bm{W}_0^{\star},\ldots,\bm{W}_t^{\star}\}$, $\{\eta_0^{\star},\ldots,\eta_{t-1}^{\star}\}$ and $(\eta_t^{\mr{out}})^{\star}$ given in Lemma 1.
\end{theorem} \par\vspace{5pt}
Theorem 2 gives the optimal LE and NLE structures for the SE of OAMP.  To compute $\eta_t^{\star}$ and $(\eta_t^{\mr{out}})^{\star}$ in \eqref{Eqn:V-A-opt}, we need to know the signal distribution $P_X(x)$. In practical applications, such prior information may be unavailable. To approach the optimal performance for OAMP, the EM learning framework \cite{Vila2013} or the parametric SURE approach \cite{GuoChunli2015} developed for AMP could be applicable to OAMP as well \cite{Xue2016}.
\subsection{Potential Optimality of OAMP}
Note that the de-correlated constraint on $\bm{W}_t$ and the divergence-free constraint on $\eta_t$ are restrictive. We next show that, provided that the SE in \eqref{Eqn:opt_SE} is valid, OAMP is potentially optimal when the optimal $\bm{W}_t^{\star}$, $\eta_t^{\star}$ and $(\eta_t^{\mr{out}})^{\star}$ given in Lemma 1 are used.\par
\begin{theorem}\label{Pro:V}
When the optimal $\{\bm{W}_t^{\star}\}$ and $\{\eta_t^{\star}\}$ in Lemma 1 are used, $\{v_t^2\}$ and $\{\tau_t^2\}$ are monotonically decreasing sequences. Furthermore, the stationary value of $\tau_t^2$, denoted by $\tau_{\infty}^2$, satisfies the following equation
\BE \label{Eqn:V-B-fixed}
\frac{1}{{\tau _\infty ^2 }} = \frac{1}{{\sigma ^2 }} \cdot R_{{\bf{A}}^{\rm{T}} {\bf{A}}} \left( { - \frac{1}{{\sigma ^2 }} \cdot mmse_B \left( {\tau _\infty ^2 } \right)} \right),
\EE
where $R_{\bm{A}^{\mr{T}}\mb{A}}$ denotes the $R$-transform \cite[pp.~48]{Tulino2004} w.r.t. the eigenvalue distribution of $\bm{A}^{\mr{T}}\bm{A}$.
\end{theorem}\par
\begin{IEEEproof}
See Appendix~\ref{App:IV}.
\end{IEEEproof}
Eqn.~\eqref{Eqn:V-B-fixed} is consistent with the fixed-point equation characterization of the MMSE performance for \eqref{Eqn:Intro} (with $\bm{A}$ being unitarily-invariant) via the replica method \cite[Eqn.~(17)]{Tulino2013}\cite[Eqn.~(30)]{Cakmak2014}. This implies that OAMP can potentially achieve the optimal MSE performance. We can see that the de-correlated and divergence-free constraints on LE and NLE, though restrictive, do not affect the potential optimality of OAMP. 
\section{Numerical Study}\label{Sec:V}
The following setups are assumed unless otherwise stated. The optimal $\bm{W}_t^{\star}$, $\eta_t^{\star}$ and $(\eta_t^{\mr{out}})^{\star}$ given in Lemma 1 are adopted for OAMP. Furthermore, the approximation $mmse_B(\tau_t^2)\approx\sum_{j=1}^N\mr{var}\big\{ x_j|r_j^t \big\}/N$ is used for \eqref{Eqn:V-A-opt-e}. Following \cite{Vila2014}, we define $\mr{SNR}\equiv\mr{E}\left\{\|\bm{Ax}\|^2\right\}/\mr{E}\left\{\|\bm{n}\right\|^2\}$.
\subsection{IID Gaussian Matrix}
\begin{figure}[htbp]
\centering
   \includegraphics[width=.5\textwidth]{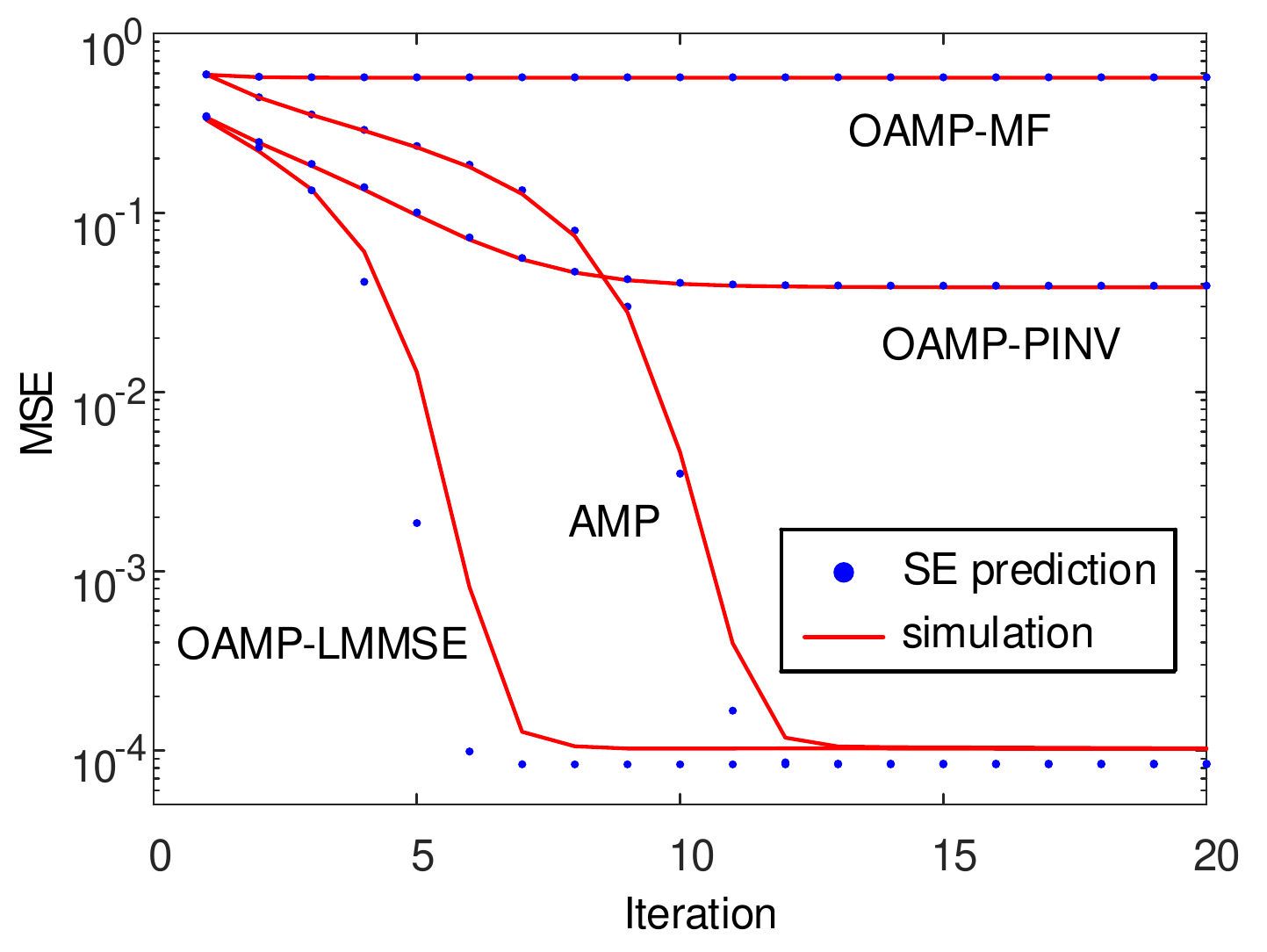}
  \caption{Simulated and predicted MSEs for OAMP with an IID Gaussian matrix and BPSK signals. $N = 8192$. $M=5324 (\approx 0.65N)$. SNR = 14 dB. The simulated MSEs are averaged over 100 realizations. }\label{Fig:Gaussian}
\end{figure}

We start from an IID Gaussian matrix where $A_{i,j}\sim\mathcal{N}(0,1/M)$. Fig.~\ref{Fig:Gaussian} compares simulated MSE with SE prediction for OAMP and AMP. We first assume that the entries of $\bm{x}$ are independently BPSK modulated, so $\bm{x}$ is not sparse. This is a typical detection problem in massive MIMO applications. Fig.~\ref{Fig:Gaussian} compares simulated MSEs with SE prediction for OAMP and AMP. In Fig.~\ref{Fig:Gaussian}, OAMP-MF, OAMP-PINV and OAMP-LMMSE refer to, respectively, OAMP algorithms with the MF, PINV and LMMSE estimators given in \eqref{Eqn:OAMP_LEs} and the normalization in \eqref{Eqn:OAMP_W_decorr}. The asymptotic SE formula in \eqref{Eqn:opt_SE_asy} becomes, respectively,
\BS \label{Eqn:Numerical_SEs}
\begin{align}
\Phi _t^{{\rm{MF}}} \left( {v_t^2 } \right) &= \frac{N}{M} \cdot v_t^2  + \sigma ^2 , \label{Eqn:Numerical_SE_a}\\
\Phi _t^{{\rm{PINV}}} \left( {v_t^2 } \right) &=
\begin{cases}
\frac{{N - M}}{M} \cdot v_t^2  + \frac{N}{{N - M}} \cdot \sigma ^2  & \text{if } M<N\label{Eqn:Numerical_SE_b}\\
\frac{M}{M-N}\cdot\sigma^2 &\text{if } M>N
\end{cases}\\
\Phi _t^{{\rm{LMMSE}}} \left( {v_t^2 } \right) &= \frac{{\sigma ^2  + c \cdot v_t^2  + \sqrt {\left( {\sigma ^2  + c \cdot v_t^2 } \right)^2  + 4\sigma ^2 v_t^2 } }}{2}, \label{Eqn:Numerical_SE_c}
\end{align}
\ES
where $c\equiv(N-M)/M$. Comparing \eqref{Eqn:Numerical_SE_a} and \eqref{Eqn:Numerical_SE_b}, we see that OAMP-PINV has better interference cancellation property than OAMP-MF (but less robust to noise). This is consistent with the observation in Fig.~\ref{Fig:Gaussian} (which represents a high SNR scenario) that OAMP-PINV can outperform OAMP-MF.\par
From Fig.~\ref{Fig:Gaussian}, we observe good agreement between the simulated and predicted MSE for all curves. Furthermore, we see that AMP has the same convergent value as OAMP-LMMSE for IID Gaussian matrices, while the latter converges faster. Following the approach in \cite{Ma2015b}, we can prove this observation but the details are omitted due to space limitation. 
\subsection{General Unitarily-invariant Matrix}
We next turn our attention to more general sensing matrices. Following \cite{Vila2014}, let $\bm{A}=\bm{V\Sigma U}^{\mr{T}}$, where $\bm{V}$ and $\bm{U}$ are independent Haar-distributed matrices (or isotropically random orthogonal matrices \cite{Tulino2004}). The nonzero singular values are set to be \cite{Vila2014} $\lambda_i/\lambda_{i+1}=\kappa^{1/M}$ for $i = 1,\ldots,M-1$, and $\sum_{j=1}^M \lambda_i=N$  Here, $\kappa\ge1$ is the condition number of $\bm{A}$. We consider sparse signals, generated according to a Bernoulli-Gaussian distribution:
\BE
P_X(x)=\rho\cdot\mathcal{N}(x;0,\rho^{-1})+(1-\rho)\cdot\delta(x),
\EE
where $\rho\in(0,1]$ is s sparsity level and $\delta(\cdot)$ is the Dirac delta function.
\par
\begin{figure}[htbp]
\centering
  \includegraphics[width=.5\textwidth]{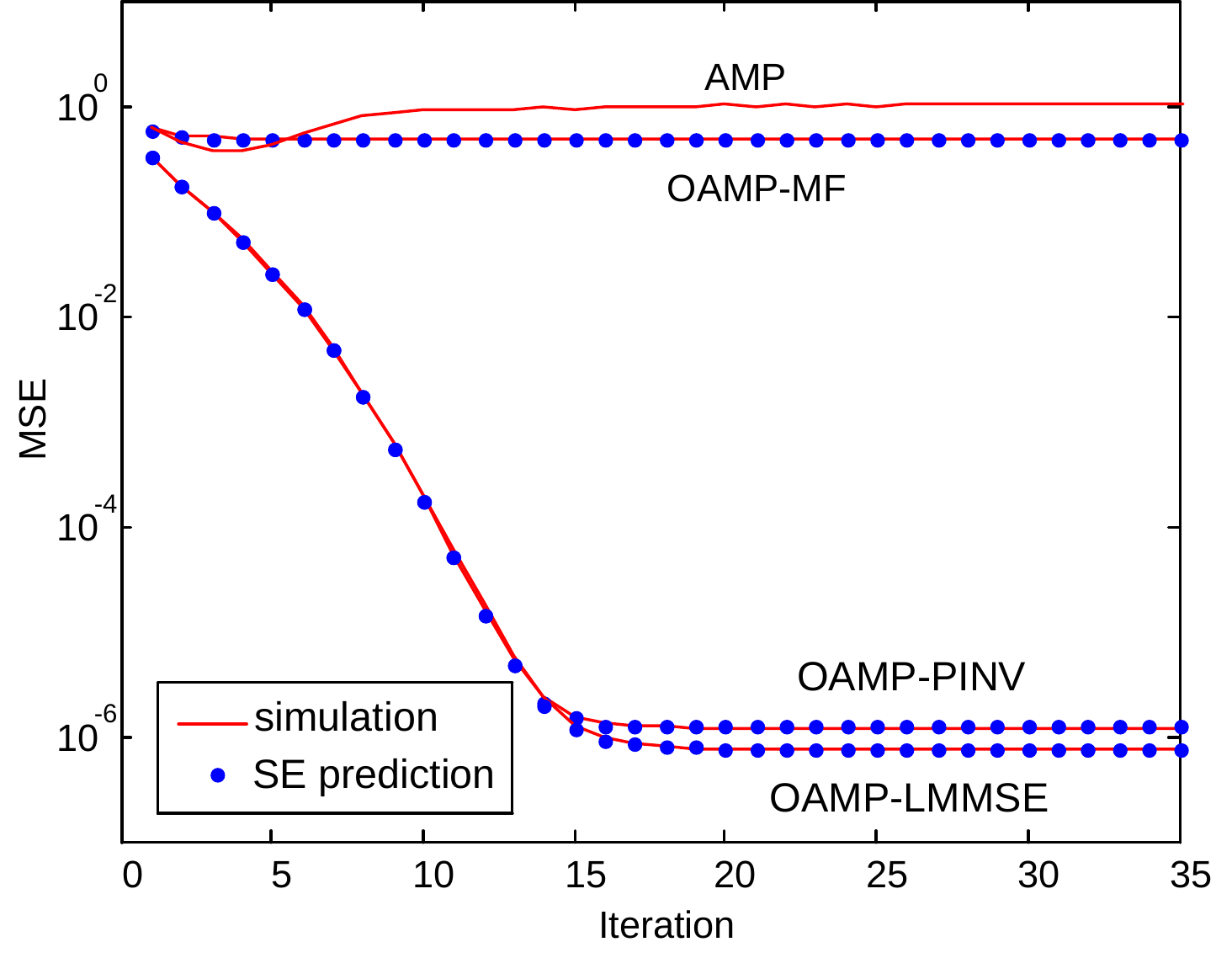}
  \caption{Simulated and predicted MSEs for OAMP with general unitarily invariant matrices. $\rho=0.2$. $N = 4000.$ $M = 2000.$ The condition number $\kappa$ is 5. SNR = $60$ dB. The simulated MSEs are averaged over 100 realizations.}\label{Fig:UDV_SE}
\end{figure}
Fig.~\ref{Fig:UDV_SE} shows the simulated and predicted MSEs for OAMP for the above ill-conditioned sensing matrix. The SE of OAMP is based on the empirical form in \eqref{Eqn:opt_SE_exp} as $\{\lambda_i\}$ are fixed in this example. We can make the following observations.
\begin{itemize}
\item The performances of AMP and OAMP-MF deteriorate in this case. The SE prediction for AMP is not shown in Fig.~\ref{Fig:UDV_SE} since it is noticeably different from the simulation result. (See Fig.~\ref{Fig:SE-div} for a similar issue.)
\item The performance of OAMP is strongly affected by the LE structure. OAMP-PINV and OAMP-LMMSE significantly outperform OAMP-MF.
\item The most interesting point is that the SE in \eqref{Eqn:opt_SE} can accurately predict the OAMP simulation results for all the LE structures in Fig.~\ref{Fig:UDV_SE}. We observed in simulations that such good agreement also holds for LEs beyond the three options shown in Fig.~\ref{Fig:UDV_SE}.
\end{itemize}\par
\begin{figure}[htbp]
\centering
  \includegraphics[width=.5\textwidth]{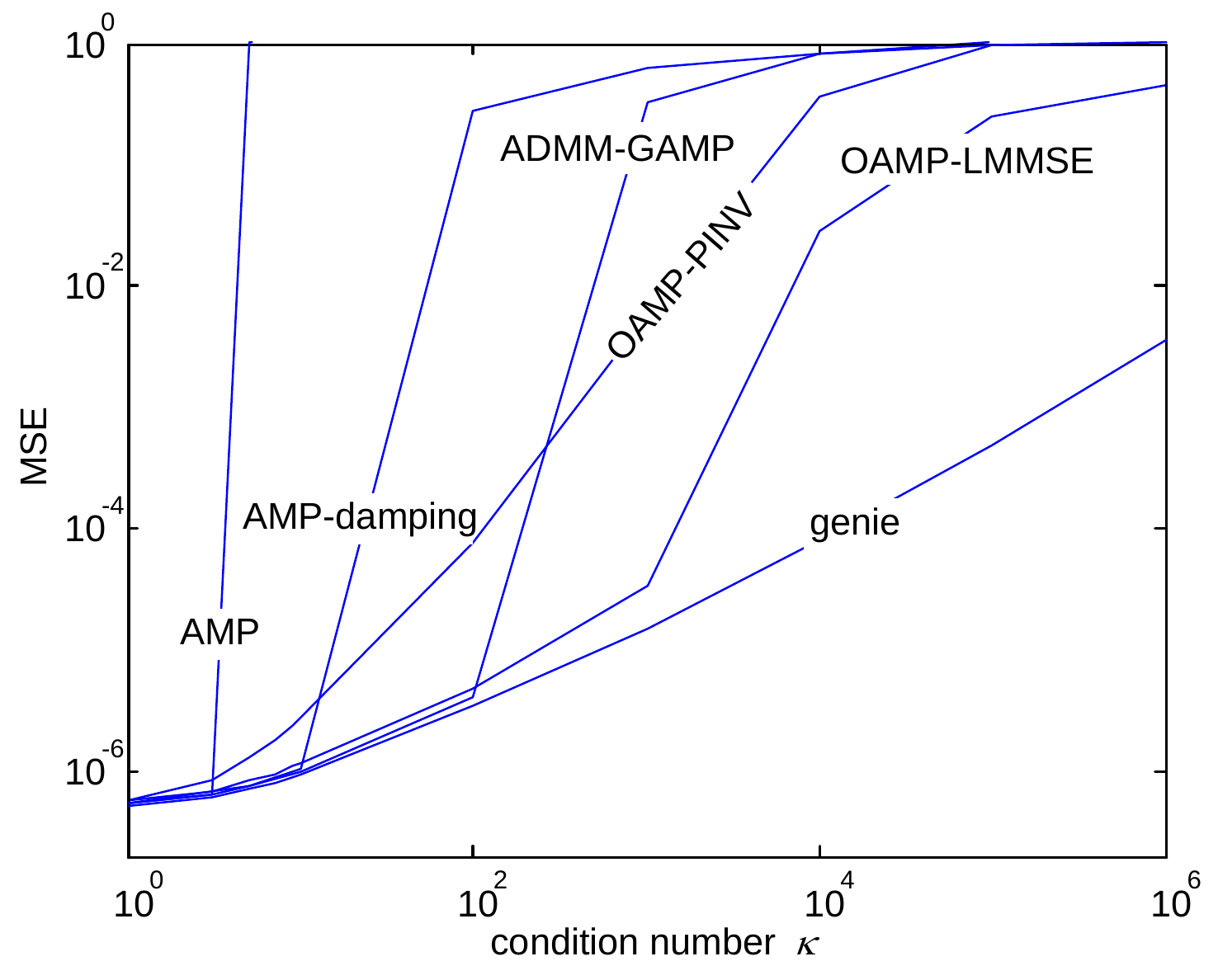}
  \caption{Comparison of OAMP and AMP for general unitarily invariant matrices. $\rho=0.2$. $N = 500.$ $M = 250.$ SNR = $60$ dB. The number of iteration for OAMP is $50$. The number of iterations for AMP and AMP-damping are $1000$. For ADMM-GAMP, both the number of inner and outer iterations are set to be 50, and the damping parameter is selected to be 1. The simulated MSEs are averaged over $100$ realizations. The MSEs above $1$ are clipped [13].}\label{Fig:UDV}
\end{figure}

Fig.~\ref{Fig:UDV} compares the MSE performances of AMP, OAMP and genie-aided MMSE (where the positions of the non-zero entries are known) as the condition number of $\bm{A}$ varies. AMP with adaptive damping (AMP-damping) \cite{Vila2014} (based on the Matlab code released by its authors\footnote{Available at http://sourceforge.net/projects/gampmatlab/} and the parameters used in \cite[Fig.~1]{Vila2014}) and GAMP-ADMM \cite{Rangan2015} are also shown. From Fig.~\ref{Fig:UDV}, we can see that the performance of OAMP-LMMSE is significantly better than those of AMP, AMP-damping and ADMM-GAMP for highly ill-conditioned scenarios. (ADMM-GAMP slightly outperforms OAMP-LMMSE for $\kappa\le100$ since the former involves more iterations in this example.) OAMP-PINV has worse performance than AMP when $\kappa\ge10$ but performs reasonably well for large $\kappa$. OAMP-MF does not work well and thus not included.\par
 For the schemes shown in Fig.~\ref{Fig:UDV}, AMP have the lowest complexity. OAMP-PINV requires one additional matrix inversion, but it can be pre-computed as it remains unchanged during the iterations. Both OAMP-LMMSE and ADMM-GAMP require matrix inversions in each iteration. As pointed out in \cite{Rangan2015}, it may be possible to replace the matrix inversion in ADMM-GAMP using an iterative method such as conjugate gradient \cite{Van2003}. Similar approximation should be possible for OAMP as well.
\subsection{Partial Orthogonal Matrix}
In the examples used above, matrix inversion is involved for $\hat{\bm{W}}^{\mr{PINV}}$ and $\hat{\bm{W}}^{\mr{LMMSE}}$ in \eqref{Eqn:OAMP_PINV} and \eqref{Eqn:OAMP_LMMSE}, so their complexity per iteration can be higher than that of AMP. (Note that the overall complexity also depends on the convergence speed, for which AMP and OAMP behave differently as seen in Fig. ~\ref{Fig:UDV}.) In the following, we will consider partial orthogonal matrices characterized by $\bm{AA}^{\mr{T}} =N/M\cdot\bm{I}$ (here $N/M$ is a normalization constant). Then inversion operation is not necessary. For example, in this case $\hat{\bm{W}}^{\mr{LMMSE}}$ is given by
\BS
\begin{align}
\hat{\bm{W}}^{\mr{LMMSE}}& = v_t^2\bm{A}^{\mr{T}}\left(v_t^2\bm{A}\bm{A}^{\mr{T}}+\sigma^2\bm{I}\right)^{-1}\\
&=\frac{v_t^2}{N/M\cdot v_t^2+\sigma^2}\cdot \bm{A}^{\mr{T}}.
\end{align}
\ES
Therefore, the complexity of OAMP-LMMSE is the same as AMP.\par
Unitarily invariant matrices with the partial orthogonality constraint becomes partial Haar-distributed matrices (i.e., uniformly distributed among all partial orthogonal matrices). We next consider the following partial orthogonal matrix
\BE \label{Eqn:VI-orth}
{\bf{A}} = \sqrt {\frac{N}{M}} {\bf{SU}}^{\rm{T}} ,
\EE
where $\bm{S}$ consists of $M$  uniformly randomly selected rows of the identity matrix and $\bm{U}$ is an Haar-distributed orthogonal matrix. We will also consider deterministic orthogonal matrices, which are important in compressed sensing and found applications in, e.g., MRI \cite{Lustig2008}. For a partial orthogonal $\bm{A}$, the three approaches in Fig.~\ref{Fig:Gaussian}, i.e., OAMP-MF, OAMP-PINV and OAMP-LMMSE, become identical. The related complexity is the same as AMP. In this case, the SE equation in \eqref{Eqn:opt_SE_exp} becomes
\BE
\Phi _t \left( {v_t^2 } \right) = \frac{{N - M}}{M} \cdot v_t^2 + \sigma ^2.
\EE\par
\begin{figure}[htbp]
\centering
  \includegraphics[width=.5\textwidth]{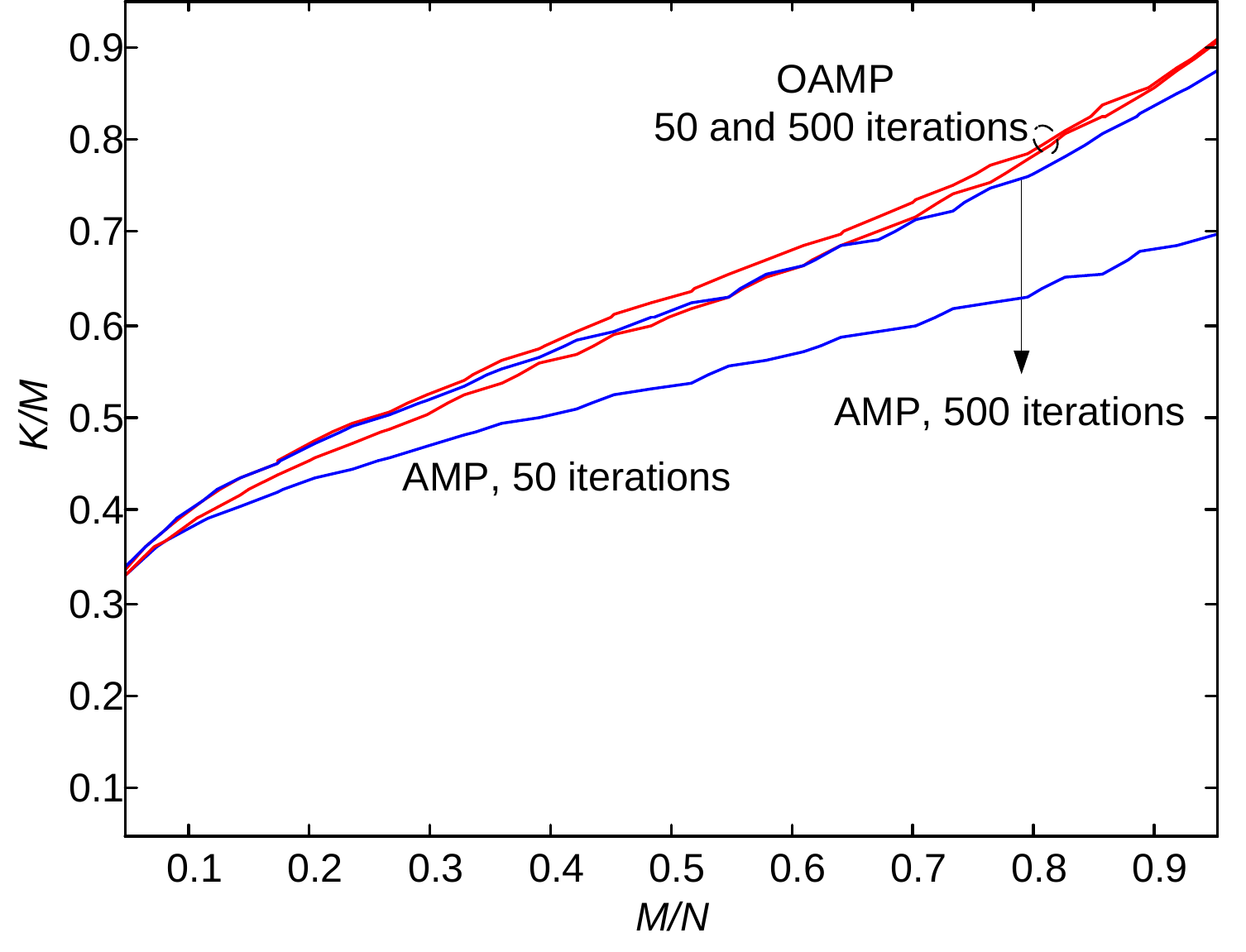}
  \caption{Noiseless empirical phase transition curves for Bernoulli-Gaussian signals with a partial DCT matrix. $N = 8192.$ The simulated MSEs are averaged over $100$ realizations. Other settings follow those of \cite[Fig.~3]{Vila2013}. Here, $K\approx N\cdot\rho$ is the average number of nonzero components in $\bm{x}$. }\label{Fig:PT}
\end{figure}

Fig. ~\ref{Fig:PT} compares OAMP with AMP in recovering Bernoulli-Gaussian signals with a partial DCT matrix. Following \cite{Vila2013}, we will use the empirical phase transition curve (PTC) to characterize the sparsity-undersampling tradeoff. A recovery algorithm ``succeeds" with high probability below the PTC and ``fails" above it. The empirical PTCs are generated according to \cite[Section IV-A]{Vila2013}. We see that OAMP considerably outperforms AMP when both algorithms are fixed to $50$ iterations. Even when the number of iterations of AMP is increased to $500$, OAMP still slightly outperforms AMP at relatively high sparsity levels.\par
\begin{figure}[htbp]
\centering
  \includegraphics[width=.5\textwidth]{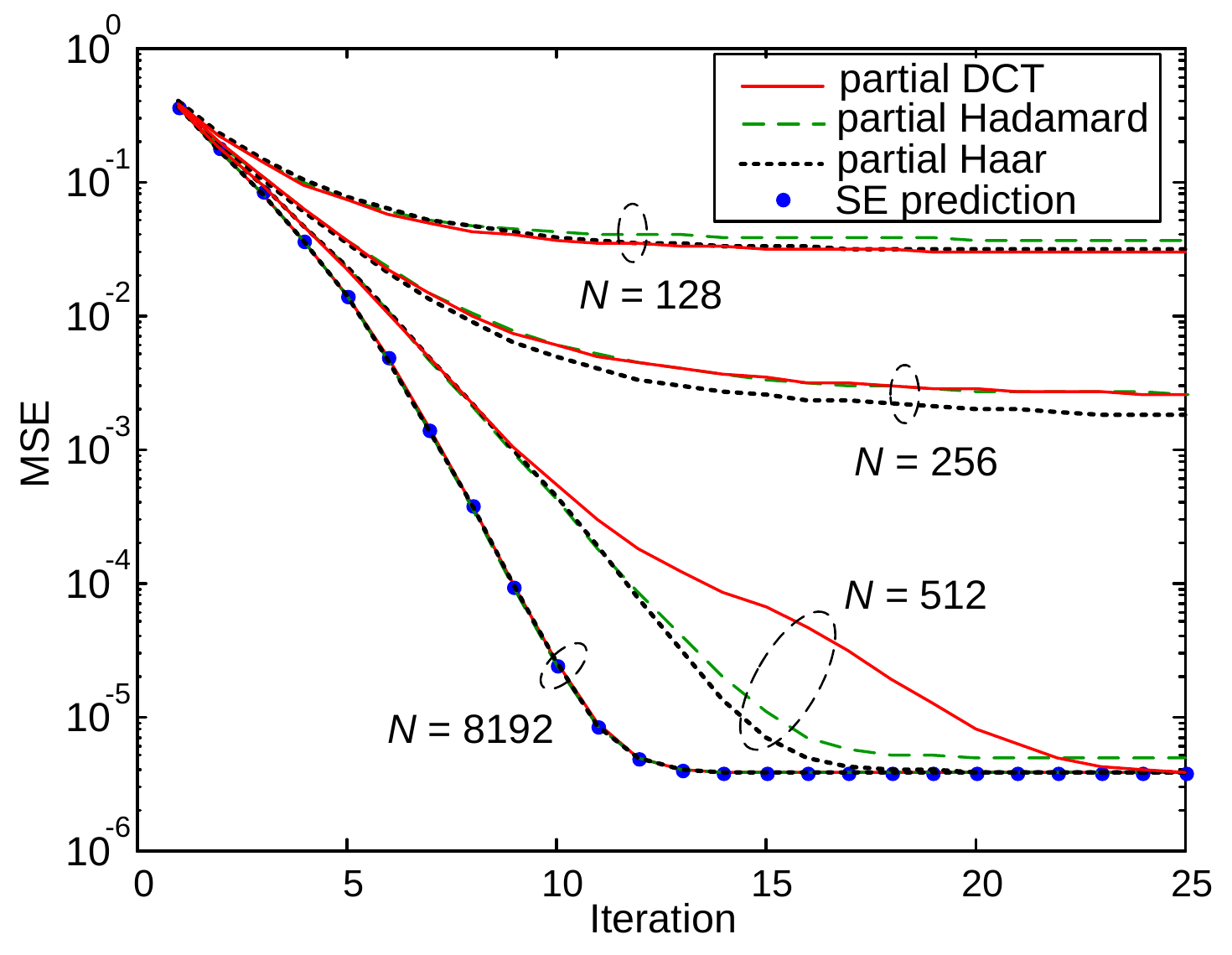}
  \caption{Simulated and predicted MSEs for OAMP with partial orthogonal matrices. $\rho=0.1$. $M=\text{round}(0.35\,N)$. SNR = 50 dB. The simulated MSEs are averaged over 2000 realizations. }\label{Fig:length}
\end{figure}

Fig.~\ref{Fig:length} shows the accuracy of SE for OAMP with partial orthogonal matrices. Three matrices are considered: a partial Haar matrix, a partial DCT matrix and a partial Hadamard matrix. From Fig.~\ref{Fig:length}, we see that the simulated MSE performances agree well with state evolution predictions for all the three types of partial orthogonal matrices when $N$ is sufficiently large ($N = 8192$ in this case). It should be noted that, when $M/N$ is larger, a smaller $N$ will suffice to guarantee good agreement between simulation and SE prediction.
\par
\begin{figure}[htbp]
\centering
  \includegraphics[width=.5\textwidth]{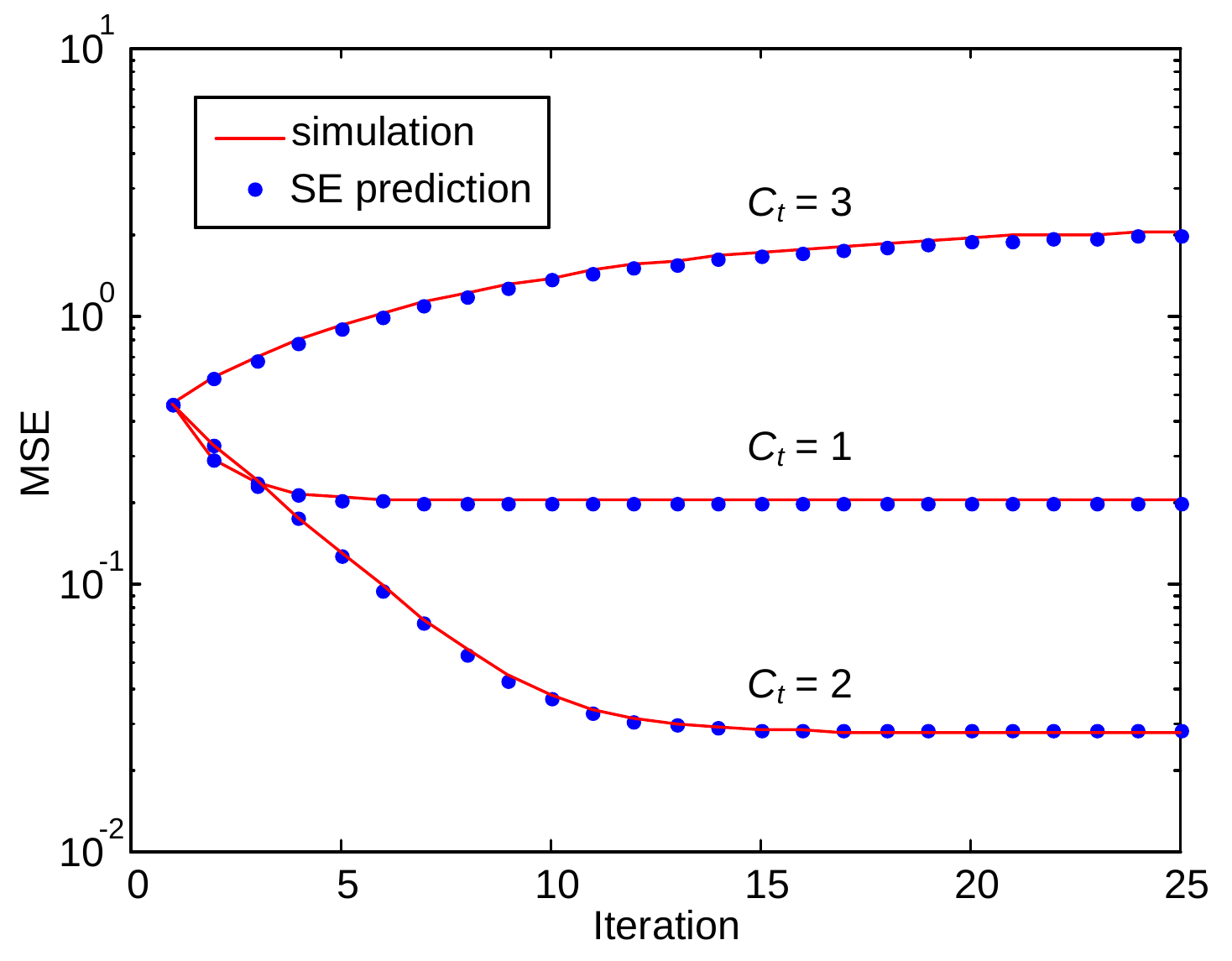}
  \caption{Simulated and predicted MSEs for OAMP with the soft-thresholding function. The threshold is set to be $\gamma_t=\tau_t$. A partial DCT matrix is used. $\rho=0.1$. $N = 8192.$ $M = 2867 (\approx0.35N)$. The simulated MSEs are averaged over $1000$ realizations.}\label{Fig:threshold}
\end{figure}

The NLEs used in Figs.~\ref{Fig:Gaussian}-\ref{Fig:length} are based on the optimized structure given in Lemma 1. Fig.~\ref{Fig:threshold} shows the OAMP SE accuracy with the following soft-thresholding function \cite{Donoho1995}:
\BE \label{Eqn:threshold}
\hat \eta_t \left( {r^t } \right) = \max \left( {\left| {r^t } \right| - \gamma _t ,0} \right) \cdot {\mathop{\rm sign}} \left( {r^t } \right),
\EE
where $\gamma_t\ge0$ is a threshold and $\mr{sign}({r}^t)$ is the sign of $r^t$. According to \eqref{Eqn:III-B-OAMP-DF}, the divergence-free function $\eta_t$ is constructed as
\BE \label{Eqn:threshold-DF}
\eta _t \left( {{\bf{r}}^t } \right) = C_t  \cdot \Bigg( {\hat \eta_t \left( {{\bf{r}}^t } \right) - \bigg( {\frac{1}{N}\sum\limits_{j = 1}^N {\mathbb{I}\left( {| {r_j^t } | > \gamma _t } \right)} } \bigg) \cdot {\bf{r}}^t } \Bigg),
\EE
where $\mathbb{I}(\cdot)$ is the indicator function. Further, we set $\eta_t^{\mr{out}}=\hat{\eta}_t$ for simplicity. The function in \eqref{Eqn:threshold} is not optimal under the MMSE sense in Lemma 1. However, it is near minimax for sparse signals \cite{Donoho13} and widely studied in compressed sensing. The optimal $C_t$ is different from that given in Lemma 1 in this case. We will not discuss details in optimizating $C_t$ here. Rather, to demonstrate the accuracy of SE, three arbitrarily chosen values for $C_t$ are used in Fig.~\ref{Fig:threshold}. We see that simulation and SE predictions agree well for all cases. In particular, when $C_t = 3$, SE is able to predict the OAMP behavior even when iterative processing leads to worse MSE performance.\par

\section{Conclusions}\label{Sec:VII}
AMP performs excellently for IID Gaussian transform matrices. The performance of AMP can be characterized by SE in this case. However, for other matrix ensembles, the SE for AMP is not directly applicable and its performance is not warranted. \par
In this paper, we proposed an OAMP algorithm based on a de-correlated LE and a divergence-free NLE. Our numerical results indicate that OAMP could be characterized by SE for general unitarily-invariant matrices with much relaxed requirements on the eigenvalue distribution and LE structure. This makes OAMP suitable for a wider range of applications than AMP, especially for applications with ill-conditioned transform matrices and partial orthogonal matrices. We also derived the optimal structures for OAMP and showed that the corresponding SE fixed point potentially coincides with that of the Bayes-optimal performance obtained by the replica method.
\section{Acknowledgement}
The authors would like to thank Dr. Ulugbek Kamilov and Prof. Phil Schniter for generously sharing their Matlab code for ADMM-GAMP.
\appendices
\section{Proof of Proposition~\ref{Pro:IV}}\label{App:I}
It is seen from \eqref{Eqn:OAMP_error_b} that $\bm{q}^t$ generated by the NLE is generally correlated with $\bm{x}$, which may lead to the correlation between $\bm{x}$ and $\bm{h}^t$. We will see below that a de-correlated LE can suppress this correlation.\par
From $\bm{A}=\bm{V\Sigma U}^{\mr{T}}$, $\bm{W}_t=\bm{U}\bm{G}_t\bm{V}^{\mr{T}}$ and $\bm{B}=\bm{I}-\bm{W}_t\bm{A}=\bm{U}(\bm{I}-\bm{G}_t\bm{\Sigma })\bm{U}^{\mr{T}}$, so
\BE
\underset{\bm{U}}{\mr{E}}\left\{ (\bm{B}_t)_{i,j} \right\} = \sum\limits_{m = 1}^N {{\rm{E}}\left\{ {U_{i,m} U_{j,m} } \right\} \cdot (1-g_m\lambda_m) },
\EE
where $g_m$ and $\lambda_m$ denote the $(m,m)$th diagonal entries of $\bm{G}_t$ and $\bm{\Sigma}$, respectively. (We define $g_m=\lambda_m=0$ for $m=M+1,\ldots,N$). For a Haar distributed matrix $\bm{U}$, we have \cite[Lemma 1.1 and Proposition 1.2]{Hiai1999}
\BE \label{Eqn:AppII-6}
\mr{E}\{U_{i,m}U_{j,m}\} =
\begin{cases}
0 & \text{if } i\neq j, \\
N^{-1} & \text{if } i = j.
\end{cases}
\EE
Therefore,
\BE \label{Eqn:App1_tr}
\underset{\bm{U}}{\mr{E}}\left\{ (\bm{B}_t)_{i,j} \right\} = \begin{cases}
0 & \text{if } i\neq j, \\
N^{-1}\mr{tr}(\bm{B}_t) & \text{if } i = j.
\end{cases}
\EE
From the discussions in Section \ref{Sec:III-A}, when $\bm{W}_t$ is de-correlated, $\mr{tr}(\bm{B}_t)=\mr{tr}(\bm{I-W}_t\bm{A})=0$. Together with \eqref{Eqn:App1_tr}, this further implies $\mr{E}\{\bm{B}_t\} = \mathbf{0}$.\par
From Assumption 1, $\bm{q}^t$ is independent of $\bm{A}$ (and so $\bm{B}_t$). Then,
\BS
\begin{align}
\mr{E}\{\bm{h}^t\} & = \mr{E}\{ \bm{B}_t\bm{q}^t \} + \mr{E}\{\bm{W}_t\bm{n}\}\\
&= \mr{E}\{ \bm{B}_t \} \mr{E}\{\bm{q}^t\} + \mr{E}\{\bm{W}_t\} \mr{E}\{\bm{n}\}\\
&=\mathbf{0}.
\end{align}
\ES
From \eqref{Eqn:OAMP_error_a}, to prove $\bm{x}$ is uncorrelated with $\bm{h}^t$, we only need to prove $\bm{x}$ is uncorrelated with $\bm{B}_t\bm{q}^t$ since $\bm{W}_t\bm{n}$ is independent of $\bm{x}$. This can be verified as
\BE
\mr{E}\left\{ \bm{B}_t \bm{q}^t \bm{x}^{\mr{T}} \right\} = \bm{E}\{ \bm{B}_t \} \mr{E}\{ \bm{q}^t\bm{x}^{\mr{T}} \} = \mathbf{0}.
\EE
\par
Following similar procedures, we can also verify that (i) the entries in $\bm{h}^t$ are uncorrelated, and (ii) the entries of $\bm{h}^t$ have identical variances. We omit the details here.

\section{Proof of Lemma 1}\label{App:II}
\subsection{Optimality of $\bm{W}_t^{\star}$}
We can rewrite $\Phi_t(v_t^2)$ in \eqref{Eqn:opt_SE_exp} as
\BE \label{Eqn:AppIII-1}
\Phi_t \left( {v_t^2 } \right) = \left(  \frac{ \frac{1}{N} \sum_{i=1}^N \hat{g}_i^2(v_t^2\lambda_i^2+\sigma^2)}{ \left( \frac{1}{N} \sum_{i=1}^N \hat{g}_i\lambda_i\right)^2 }  \right)
 - v_t^2 .
\EE
We now prove that $\bm{W}_t^{\star}$ in Lemma 1 is optimal for \eqref{Eqn:AppIII-1}. To this end, define $a_i \equiv \hat{g}_i  \sqrt { { v_t^2 \lambda_i^2   + \sigma ^2 } }$, $b_i \equiv \lambda_i/\sqrt { {   v_t^2\lambda_i^2 + \sigma ^2 } }$. Applying the Cauchy-Schwarz inequality
\BE
\frac{\frac{1}{N}\sum_{i=1}^Na_i^2}{ \left(\frac{1}{N}\sum_{i=1}^Na_ib_i\right)^2}\ge\left(\frac{1}{N}\sum_{i=1}^N b_i^2\right)^{-1}
\EE
leads to
\BE \label{Eqn:AppIII-2}
\frac{ \frac{1}{N} \sum_{i=1}^N \hat{g}_i^2(v_t^2\lambda_i^2+\sigma^2)}{ \left( \frac{1}{N} \sum_{i=1}^N \hat{g}_i\lambda_i\right)^2 } \ge \left( \frac{1}{N} \sum_{i=1}^N \frac{\lambda_i^2}{v_t^2\lambda_i^2+\sigma^2} \right)^{-1},
\EE
where the right hand side of \eqref{Eqn:AppIII-2} is invariant to $\{\hat{g}_i\}$. The minimum in \eqref{Eqn:AppIII-2} is reached when
\BE \label{Eqn:AppII-Cauchy}
\hat{g}_i^{\star} \sqrt { {v_t^2 \lambda_i^2  + \sigma ^2 } } = C \sqrt {\frac{\lambda_i^2}{ {v_t^2  \lambda_i^2  + \sigma ^2} } },
\EE
where $C$ is an arbitrary constant. From \eqref{Eqn:AppII-Cauchy},
\BE \label{Eqn:AppIII-3}
\hat{g}_i^{\star}  = C \frac{\lambda_i}{{v_t^2   \lambda_i^2  + \sigma ^2 }}.
\EE
Recall that $\{\lambda_i\}$ are the singular values of $\bm{A}$. Setting $C=v_t^2$, we can see that $\{\hat{g}_i^{\star}\}$ obtained from \eqref{Eqn:AppIII-3} are the singular values of $\hat{\bm{W}}_t^{\mr{LMMSE}}\equiv v_t^2\bm{A}^{\mr{T}}(v_t^2\bm{AA}^{\mr{T}}+\sigma^2\bm{I})^{-1}$ in \eqref{Eqn:OAMP_LMMSE}. Therefore the optimal $\bm{W}_t^{\star}$ can be obtained by substituting $\hat{\bm{W}}_t^{\star}=\hat{\bm{W}}_t^{\mr{LMMSE}}$ into \eqref{Eqn:OAMP_W_decorr}:
\BE
\bm{W}_t^{\star}=\frac{N}{\mr{tr}(\hat{\bm{W}}_t^{\mr{LMMSE}}\bm{A})}\hat{\bm{W}}_t^{\mr{LMMSE}}.
\EE
\subsection{Optimality of $\eta_t^{\star}$}
The SE equation in \eqref{Eqn:opt_SE_psi} are obtained based on the following signal model
\BE
R^t  = X + \tau _t Z.
\EE
The following identity is from \cite[Eqn.~(123)]{Rangan2011}
\BE \label{Eqn:AppIII-5}
\frac{{{\rm{d}} {\eta _t^{{\rm{MMSE}}} }   }}{{{\rm{d}}R^t }} = \frac{1}{{\tau _t^2 }} \cdot {\rm{var}}\left\{ {X|R^t } \right\},
\EE
where ${\eta _t^{{\rm{MMSE}}} }   \equiv {\rm{E}}\left\{ {X|R^t } \right\}$ (see \eqref{Eqn:V-A-opt-b}). Using \eqref{Eqn:AppIII-5} and noting $mmse_B(\tau_t^2)=\mr{E}\{\mr{var}\{X|R^t\}\}$, we can verify that $\eta_t^{\star}$ in \eqref{Eqn:V-A-opt-c} is a divergence-free function (see \eqref{Eqn:III-A-DF-fun}).\par
Lemma~\ref{Lem:4} below is the key to prove the optimality of $\eta_t^{\star}$.
\begin{lemma}\label{Lem:4}
The following holds for any divergence-free function $\eta_t$
\BE
{\rm{E}}\left\{ {\eta _t  \cdot \left( \eta_t^{\mr{MMSE}}   - \eta _t^{\star}  \right)} \right\} = 0.
\EE
\end{lemma}
\begin{IEEEproof} We can rewrite \eqref{Eqn:V-A-opt-c} as
\BE  \label{Eqn:AppIII-7}
\eta _t^{\star}  = C_t^{\star}  \cdot \eta_t^{\mr{MMSE}}   + \left( {1 - C_t^{\star} } \right) \cdot R^t .
\EE
First,
\BS \label{Eqn:AppIII-8}
\begin{align}
\eta_t^{\mr{MMSE}} - \eta _t^{\star} & = \eta_t^{\mr{MMSE}}   - \left[ {C_t^*  \cdot \eta_t^{\mr{MMSE}}   + \left( {1 - C_t^* } \right) \cdot R^t } \right]\\
& = \left( {1 - C_t^{\star}} \right) \cdot \left( {\eta_t^{\mr{MMSE}}  - R^t } \right).
\end{align}
\ES
Therefore, to prove Lemma~\ref{Lem:4}, we only need to prove
\BE \label{Eqn:AppIII-9}
{\rm{E}}\left\{ {\eta _t  \cdot \left( {\eta_t^{\mr{MMSE}}   - R^t } \right)} \right\} = 0.
\EE
Substituting $R^t=X+\tau_t Z$ into \eqref{Eqn:AppIII-9} yields
\BE \label{Eqn:AppIII-10}
{\rm{E}}\left\{ {\eta _t  \cdot \left( {\eta_t^{\mr{MMSE}} - X - \tau _t Z} \right)} \right\} = 0.
\EE
Since $\eta_t$ is a divergence-free function of $R^t$, we have the following from \eqref{Eqn:III-A-orth2}
\BE \label{Eqn:AppIII-11}
{\rm{E}}\left\{ {\eta _t  \cdot Z} \right\} = 0.
\EE
Substituting \eqref{Eqn:AppIII-11} into \eqref{Eqn:AppIII-10}, proving Lemma~\ref{Lem:4} becomes proving
\BE \label{Eqn:AppIII-12}
{\rm{E}}\left\{ {\eta _t  \cdot \left( {\eta_t^{\mr{MMSE}}   - X} \right)} \right\} = 0.
\EE
Note that $\eta_t$ and $\eta_t^{\mr{MMSE}}$ are deterministic functions of $R^t$. Then, conditional on $R^t$, we have
\BS \label{Eqn:AppIII-13}
\begin{align}
{\rm{E}}\left\{ {\eta _t  \cdot \left( {\eta_t^{\mr{MMSE}}  - X} \right)|R^t } \right\} &= \eta _t  \cdot \left( {\eta_t^{\mr{MMSE}}   - {\rm{E}}\left\{ {X|R^t } \right\}} \right) \\
&= \eta _t  \cdot \left( {\eta_t^{\mr{MMSE}}  -\eta_t^{\mr{MMSE}}  } \right)\label{Eqn:AppIII-13-b}\\
& = 0,
\end{align}
\ES
where \eqref{Eqn:AppIII-13-b} is from the definition of $\eta_t^{\mr{MMSE}}$ in \eqref{Eqn:V-A-opt-b}. Therefore,
\BE \label{Eqn:AppIII-14}
{\rm{E}}\left\{ {\eta _t  \cdot \left( {\eta_t^{\mr{MMSE}}   - X} \right)} \right\}{\rm{ = }}\mathop {\rm{E}}\limits_{R^t } \left\{ {{\rm{E}}\left\{ {\eta _t  \cdot \left( {\eta_t^{\mr{MMSE}}   - X} \right)|R^t } \right\}} \right\} = 0,
\EE
which concludes the proof of Lemma~\ref{Lem:4}.
\end{IEEEproof}
\par
We next prove the optimality of $\eta_t^{\star}$ based on Lemma~\ref{Lem:4}. Again, let $\eta_t$ be an arbitrary divergence-free function of $R^t$. The estimation MSE of $\eta_t$ reads
\BS \label{Eqn:AppIII-15}
\begin{align}
\Psi _t(\tau_t^2) &\equiv {\rm{E}}\left\{ {\left( {\eta _t  - X} \right)^2 } \right\}\\
& = {\rm{E}}\left\{ {\left( {\eta _t  - \eta_t^{\mr{MMSE}}  + \eta_t^{\mr{MMSE}}  - X} \right)^2 } \right\} \label{Eqn:AppIII-15-a}\\
& = {\rm{E}}\left\{ {\left( {\eta _t  - \eta_t^{\mr{MMSE}} } \right)^2 } \right\} + {\rm{E}}\left\{ {\left( {\eta_t^{\mr{MMSE}}  - X} \right)^2 } \right\}
\label{Eqn:AppIII-15-b}\\
& = {\rm{E}}\left\{ {\left( {\eta _t  - \eta_t^{\mr{MMSE}}} \right)^2 } \right\} + mmse_B \left( {\tau _t^2 } \right),
\label{Eqn:AppIII-15-c}
\end{align}
\ES
where the cross terms in \eqref{Eqn:AppIII-15-b} disappears due to the orthogonality property of MMSE estimation \cite{Kay1993} (recall that $\eta_t^{\mr{MMSE}}$ is the scaler MMSE estimator). We see from \eqref{Eqn:AppIII-15} that finding $\eta_t$ that minimizes $\mr{E}\left\{(\eta_t-X)^2\right\}$ is equivalent to finding $\eta_t$ minimizing $\mr{E}\left\{\left(\eta_t-\eta_t^{\mr{MMSE}}\right)^2\right\}$. We can further rewrite $\mr{E}\left\{\left(\eta_t-\eta_t^{\mr{MMSE}}\right)^2\right\}$ as
\BS \label{Eqn:AppIII-16}
\begin{align}
&{\rm{E}}\left\{ {\left( {\eta _t  - \eta_t^{\mr{MMSE}} } \right)^2 } \right\} \\
&= {\rm{E}}\left\{ {\left( {\eta _t  - \eta _t^{\star}  + \eta _t^{\star}  - \eta_t^{\mr{MMSE}} } \right)^2 } \right\}\\
& = {\rm{E}}\left\{ {\left( {\eta _t  - \eta _t^{\star} } \right)^2 } \right\} + {\rm{E}}\left\{ {\left( {\eta _t^{\star}  - \eta_t^{\mr{MMSE}} } \right)^2 } \right\} \\
\nonumber &+ 2 \cdot {\rm{E}}\left\{ {\left( {\eta _t  - \eta _t^{\star} } \right)\left( {\eta _t^{\star}  - \eta_t^{\mr{MMSE}} } \right)} \right\}.
\end{align}
\ES
From Lemma~\ref{Lem:4}, we have ${\rm{E}}\left\{ {\eta _t  \cdot \left( {\eta _t^{\star}  - \eta_t^{\mr{MMSE}} } \right)} \right\} = 0$ and ${\rm{E}}\left\{ {\eta _t^{\star}  \cdot \left( {\eta _t^{\star}  - \eta_t^{\mr{MMSE}} } \right)} \right\} = 0$ (since $\eta_t^{\star}$ is itself a divergence-free function). Then, \eqref{Eqn:AppIII-16} becomes
\BS
\begin{align}
&{\rm{E}}\left\{ {\left( {\eta _t  - \eta_t^{\mr{MMSE}} } \right)^2 } \right\}\\
& = {\rm{E}}\left\{ {\left( {\eta _t  - \eta _t^{\star} } \right)^2 } \right\} + {\rm{E}}\left\{ {\left( {\eta _t^{\star}  - \eta_t^{\mr{MMSE}} } \right)^2 } \right\}. \label{Eqn:AppIII-17} \\
&\ge {\rm{E}}\left\{ {\left( {\eta _t^{\star}  - \eta_t^{\mr{MMSE}} } \right)^2 } \right\},  \label{Eqn:AppIII-18}
\end{align}
\ES
where the equality is obtained when $\eta_t=\eta_t^{\star}$, and the right hand side of \eqref{Eqn:AppIII-18} is a constant invariant of $\eta_t$. Hence, $\eta_t=\eta_t^{\star}$ minimizes ${\rm{E}}\left\{ {\left( {\eta _t  - \eta_t^{\mr{MMSE}} } \right)^2 } \right\}$ and so $\Psi_t\equiv\mr{E}\left\{(\eta_t-X)^2\right\}$. This completes the proof. 
\section{Proof of Lemma 2}\label{App:III}
\subsection{Derivation of $\Psi^{\star}$ in \eqref{Eqn:opt_SE_opt_b}}\label{App:III-A}
Using \eqref{Eqn:AppIII-7}, we have
\BS \label{Eqn:AppIV-3}
\begin{align}
&\Psi ^{\star}  \left( {\tau_t^2 } \right) \\
= &{\rm{E}}\left\{ {\left( {\eta _t^{\star}  - X} \right)^2 } \right\} \\
= &{\rm{E}}\left\{ {\left[ {C_t^{\star}  \cdot \eta_t^{\mr{MMSE}}  + \left( {1 - C_t^{\star} } \right) \cdot R^t  - X} \right]^2 } \right\} \label{Eqn:AppIV-3-a}\\
\nonumber = &\left( {C_t^{\star} } \right)^2   {\rm{E}}\left\{  \left( {\eta_t^{\mr{MMSE}}- X} \right)^2 \right\} + \left( {1 - C_t^{\star} } \right)^2   {\rm{E}}\left\{ {\left( {R^t  - X} \right)^2 } \right\} \label{Eqn:AppIV-3-b} \\
 & + 2C_t^{\star} \left( {1 - C_t^{\star} } \right){\rm{E}}\left\{ {\left( {\eta_t^{\mr{MMSE}}  - X} \right)\tau _t Z} \right\} \\
 \nonumber= &\left( {C_t^{\star} } \right)^2  \cdot mmse_B \left( {\tau _t^2 } \right) + \left( {1 - C_t^{\star} } \right)^2  \cdot \tau _t^2 \\
 & + 2C_t^{\star} \left( {1 - C_t^{\star} } \right) \cdot mmse_B\left( {\tau _t^2 } \right)   \label{Eqn:AppIV-3-c}\\
 =& \left( {\frac{1}{{mmse_B \left( {\tau _t^2 } \right)}} - \frac{1}{{\tau _t^2 }}} \right)^{ - 1} ,\label{Eqn:AppIV-3-d}
\end{align}
\ES
where \eqref{Eqn:AppIV-3-c} is from the fact that $\mr{E}\{XZ\}=0$, Stein's lemma and \eqref{Eqn:AppIII-5}, \eqref{Eqn:AppIV-3-d} from the definition of $C_t^{\star}$ in \eqref{Eqn:V-A-opt}.
\subsection{Monotonicity of $\Phi^{\star}$ and $\Psi^{\star}$} \label{App:III-B}
We first verify the monotonicity of $\Phi^{\star}$. From \eqref{Eqn:opt_SE_opt_a} and after some manipulations, we obtain
\BE \label{Eqn:AppIV-4}
\frac{{{\rm{d}}\Phi ^ {\star} }}{{{\rm{d}}v_t^2 }} = \frac{{\left( {v_t^2 } \right)^2  \cdot \frac{{{\rm{d}}mmse_A \left( {v_t^2 } \right)}}{{{\rm{d}}v_t^2 }} - \left[ {mmse_A \left( {v_t^2 } \right)} \right]^2 }}{{\left[ {v_t^2  - mmse_A \left( {v_t^2 } \right)} \right]^2 }}.
\EE
To show the monotonicity of $\Phi^{\star}$, we only need to show that
\BE \label{Eqn:AppIV-5}
\frac{{{\rm{d}}mmse_A \left( {v_t^2 } \right)}}{{{\rm{d}}v_t^2 }} \ge \left( {\frac{{mmse_A \left( {v_t^2 } \right)}}{{v_t^2 }}} \right)^2 .
\EE
The derivative of $mmse_A \left( {v_t^2 } \right)$ can be computed based on the definition below \eqref{Eqn:opt_SE_opt}. After some manipulations, the inequality in \eqref{Eqn:AppIV-5} becomes the inequality below
\BE \label{Eqn:AppIV-6}
\frac{1}{N}\sum_{i=1}^N\left( \frac{\sigma^2}{v_t^2\lambda_i^2+\sigma^2} \right)^2\ge
\left(\frac{1}{N}\sum_{i=1}^N\frac{\sigma^2}{v_t^2\lambda_i^2+\sigma^2}\right)^2,
\EE
which holds due to Jensen's inequality.\par
The monotonicity of $\Psi^{\star}$ can be proved in a similar way. Again, we only need to prove that
\BE \label{Eqn:AppIV-7}
\frac{{{\rm{d}}mmse_B \left( {\tau _t^2 } \right)}}{{{\rm{d}}\tau _t^2 }} \ge \left( {\frac{{mmse_B \left( {\tau _t^2 } \right)}}{{\tau _t^2 }}} \right)^2 .
\EE
Note that $mmse_B \left( {\tau _t^2 } \right) = {\rm{E}}\left\{ {\left[ {X - {\rm{E}}\left\{ {X|R^t  = X{\rm{ + }}\tau _t Z} \right\}} \right]^2 } \right\}$. From \cite[Proposition 9]{Guo2011}, we have
\BE \label{Eqn:AppIV-8}
\frac{{{\rm{d}}mmse_B \left( {\tau _t^2 } \right)}}{{{\rm{d}}\tau _t^2 }} =  {\frac{{{\rm{E}}\left\{ {{\rm{var}}\left\{ {X|R^t } \right\}^2 } \right\}}}{{\left(\tau _t^2\right)^2 }}}  .
\EE
Applying Jensen's inequality, we have
\BE \label{Eqn:AppIV-9}
{\rm{E}}\left\{ {{\rm{var}}\left\{ {X|R^t } \right\}^2 } \right\} \ge \left[ {{\rm{E}}\left\{ {{\rm{var}}\left\{ {X|R^t } \right\}} \right\}} \right]^2  = \left[ {mmse_B \left( {\tau _t^2 } \right)} \right]^2 ,
\EE
which, together with \eqref{Eqn:AppIV-8}, proves \eqref{Eqn:AppIV-7}. 
\section{Proof of Theorem \ref{Pro:V}} \label{App:IV}
\subsection{Monotonicity of $\{v_t^2\}$ and $\{\tau_t^2\}$}
We first show that $\{v_t^2\}$ decrease monotonically. From \eqref{Eqn:opt_SE_opt_b},
\BS
\begin{align}
\lim_{\tau^2\to\infty}\Psi^{\star}(\tau^2)&=\lim_{\tau^2\to\infty}\frac{\tau^2\cdot mmse_B(\tau^2)}{\tau^2-mmse(\tau^2)}\\&=\lim_{\tau^2\to\infty}mmse_B(\tau^2) \label{Eqn:opt_mono_b}\\
&=\mr{E}\{X^2\}\\
&=v_0^2, \label{Eqn:opt_mono_c}
\end{align}
\ES
where \eqref{Eqn:opt_mono_c} is from the initialization of the SE. Since $\Phi^{\star}(v_0^2)<\infty$ and $\Psi^{\star}$ is a monotonically increasing function, we have $v_1^2=\Psi^{\star}\left(\Phi^{\star}(v_0^2)\right)<v_0^2$. \par
We now proceed by induction. Suppose that $v_t^2<v_{t-1}^2$. Since both $\Phi^{\star}$ and $\Psi^{\star}$ are monotonically increasing, we have $\Psi^{\star}\left(\Phi^{\star}(v_t^2)\right)<\Psi^{\star}\left(\Phi^{\star}(v_{t-1}^2)\right)$, which, together with the SE relationship $v_{t+1}^2=\Psi^{\star}\left(\Phi^{\star}(v_t^2)\right)$, leads to $v_{t+1}^2<v_{t}^2$. Hence, $\{v_t^2\}$ is a monotonically decreasing sequence.\par
The monotonicity of the sequence $\{\tau_t^2\}$ follows directly from the monotonicity of $\{v_t^2\}$, the SE $\tau_t^2=\Phi^{\star}(v_t^2)$, and the fact that $\Phi^{\star}$ is a monotonically increasing function.
\subsection{Fixed Point Equation of SE}
Similar to \eqref{Eqn:opt_SE_asy},
\BE
mmse_A \left( {v_t^2 } \right) \equiv \frac{1}{N}\sum_{i=1}^N\frac{v_t^2\cdot \sigma^2}{v_t^2 \cdot \lambda_i^2+\sigma^2}\to{\rm{E}}\left\{ {\frac{{v_t^2 \cdot \sigma ^2  }}{{v_t^2  \cdot \lambda^2  + \sigma ^2 }}} \right\},
\EE
where the expectation is w.r.t. the asymptotic eigenvalue distribution of $\bm{A}^{\mr{T}}\bm{A}$. From the definition of the $\eta$-transform in \cite[pp.~40]{Tulino2004}, we can write
\BE \label{Eqn:AppV-1}
v_t^2  \cdot \eta _{{\bf{A}}^{\rm{T}} {\bf{A}}} \left( {\frac{{v_t^2 }}{{\sigma ^2 }}} \right)={\rm{E}}\left\{ {\frac{{ v_t^2 \cdot \sigma ^2 }}{{v_t^2  \cdot \lambda^2  + \sigma ^2 }}} \right\},
\EE
where $\eta _{{\bf{A}}^{\rm{T}} {\bf{A}}}$ denotes the $\eta$-transform. For convenience, we further rewrite \eqref{Eqn:AppV-1} as
\BE \label{Eqn:AppV-2}
\gamma  \cdot \eta _{{\bf{A}}^{\rm{T}} {\bf{A}}} \left( \gamma  \right) = \frac{1}{{\sigma ^2 }} \cdot mmse_A \left( {v_t^2 } \right).
\EE
where $\gamma\equiv v_t^2/\sigma^2$. Note the following relationship between the $\eta$-transform and the $R$-transform \cite[Eqn.~(2.74)]{Tulino2004}
\BE \label{Eqn:AppV-3}
R_{{\bf{A}}^{\rm{T}} {\bf{A}}} \left( { - \gamma  \cdot \eta _{{\bf{A}}^{\rm{T}} {\bf{A}}} \left( \gamma  \right)} \right) = \frac{1}{{\gamma  \cdot \eta _{{\bf{A}}^{\rm{T}} {\bf{A}}} \left( \gamma  \right)}} - \frac{1}{\gamma }.
\EE
Substituting \eqref{Eqn:AppV-2} into \eqref{Eqn:AppV-3} yields
\BE \label{Eqn:AppV-4}
R_{{\bf{A}}^{\rm{T}} {\bf{A}}} \left( { - \frac{1}{{\sigma ^2 }}  mmse_A \left( {v_t^2 } \right)} \right) = \frac{{\sigma ^2 }}{{mmse_A \left( {v_t^2 } \right)}} - \frac{{\sigma ^2 }}{{v_t^2 }} = \sigma ^2   \frac{1}{{\tau_t^2 }},
\EE
where the second equality in \eqref{Eqn:AppV-4} is from \eqref{Eqn:opt_SE_a} and \eqref{Eqn:opt_SE_opt_a}. We can rewrite the SE equations in \eqref{Eqn:opt_SE_opt_a} and \eqref{Eqn:opt_SE_opt_b} as follows
\BS \label{Eqn:AppV-5}
\begin{align}
mmse_A \left( {v_t^2 } \right) &= \left( {\frac{1}{{\tau _t^2 }} + \frac{1}{{v_t^2 }}} \right)^{ - 1} , \\
mmse_B \left( {\tau _t^2 } \right)& = \left( {\frac{1}{{v_{t + 1}^2 }} + \frac{1}{{\tau _t^2 }}} \right)^{ - 1} .
\end{align}
\ES
At the stationary point, we have
\BE \label{Eqn:AppV-6}
mmse_A \left( {v_\infty ^2 } \right) = mmse_B \left( {\tau _\infty ^2 } \right).
\EE
Substituting \eqref{Eqn:AppV-6} into \eqref{Eqn:AppV-4}, we get the desired fixed point equation
\BE
\frac{1}{{\tau _\infty ^2 }} = \frac{1}{{\sigma ^2 }} \cdot R_{{\bf{A}}^{\rm{T}} {\bf{A}}} \left( { - \frac{1}{{\sigma ^2 }} \cdot mmse_B \left( {\tau_\infty ^2 } \right)} \right).
\EE 

\ifCLASSOPTIONcaptionsoff
  \newpage
\fi
\bibliographystyle{IEEEtran}	
\bibliography{IEEEabrv,OAMP}		

\end{document}